\documentclass{jfm}%[lineno]{jfm}
\usepackage{ulem}
\usepackage{graphicx}
\usepackage{newtxtext}
\usepackage{newtxmath}
\usepackage{natbib}
\usepackage{hyperref}
\usepackage{xcolor}
\hypersetup{
    colorlinks = true,
    urlcolor   = blue,
    citecolor  = black,
}

\newcommand{\RomanNumeralCaps}[1]
\linenumbers
\def\Ro{\mathit{Ro}}
\def\Fr{\mathit{Fr}}
\def\Pe{\mathit{Pe}}
\def\Rey{\mathit{Re}}

\def\BLUE#1{\textcolor{black}{#1}}

\title{Energy cascades in rapidly rotating and stratified turbulence within elongated domains}

\author{Adrian van Kan\aff{1}
  \corresp{\email{avankan@ens.fr}},
 \and Alexandros Alexakis \aff{1}}

\affiliation{Laboratoire de Physique de l’Ecole normale supérieure, ENS, Université PSL, CNRS, Sorbonne Université, Université de Paris, F-75005 Paris, France}

\begin{document}
\maketitle

\begin{abstract}
We study forced, rapildy rotating and stably stratified turbulence in an elongated domain using an asymptotic expansion at simultaneously low Rossby number $\Ro\ll1$ and large domain height compared to the energy injection scale, $h=H/\ell_{in}\gg1$. The resulting equations depend on the parameter $\lambda =(h \Ro )^{-1}$ and the Froude number $\Fr$. \textcolor{black}{An extensive set of direct numerical simulations (DNS) is performed to explore the parameter space $(\lambda,\Fr)$. We show that a forward energy cascade occurs in one region of this space, and a split energy cascade outside it. At weak stratification (large $\Fr$), an inverse cascade is observed for sufficiently large $\lambda$. At strong stratification (small $\Fr$) the flow becomes approximately hydrostatic and an inverse cascade is always observed. For both weak and strong stratification, we present theoretical arguments supporting the observed energy cascade phenomenology.} Our results shed light on an asymptotic region in the phase diagram of rotating and stratified turbulence, which is difficult to attain by brute-force DNS.
\end{abstract}


\section{Introduction}
\label{sec:intro}

Rotating and stratified flows abound in the universe, from distant planets and stars to Earth's atmosphere and oceans \citep{salmon1998lectures,pedlosky2013geophysical,vallis2017atmospheric}, motivating a large number of theoretical and experimental studies in the past \citep{trustrum1964rotating,maxworthy1975experiments,gibson1991laboratory,davidson2013turbulence}. Typically these flows are turbulent, since they are characterised by large values of the Reynolds number $\Rey$, defined as the ratio of inertial forces to viscous forces. Also, the P\'eclet number $\Pe$, given by the advective rate of change of temperature over the diffusive rate of change, is typically large. In a rotating system a Coriolis force arises, whose magnitude relative to the inertial force  is measured by the Rossby number $\Ro=\frac{U}{\Omega \ell}$, where $U,\ell$ are typical flow velocity and length scales, and $\Omega$ is the rotation rate. Density stratification and gravity give rise to buoyancy forces, whose strength relative to inertial forces is measured by the Froude number $\Fr=\frac{U}{N\ell}$, where $N$ is the buoyancy frequency. For $\Ro<\infty$ and/or $\Fr<\infty$, the isotropy of three-dimensional (3-D) turbulence is broken, since the rotation axis and gravity impose a direction in space. When $\Omega$ is large, i.e. in the limit $\Ro \to 0$, rotation suppresses variations of the motion along the axis of rotation and thus makes the flow quasi-two-dimensional, an effect described by the Taylor-Proudman theorem \citep{hough1897ix, proudman1916motion, taylor1917motion, greenspan1968theory}. \textcolor{black}{Similarly, when $N$ is large, vertical motions are suppressed, and quasi-horizontal layers, so-called ``pancakes'', are favoured \citep{herring1989numerical,waite2004stratified,brethouwer2007scaling}. A review of rotating and stratified flows is given in \citep{pouquet2017dual}.}

Turbulent energy transfer strongly depends on the dimension of space. In homogeneous isotropic 3-D turbulence, energy  injected at large scales is transferred, by non-linear interactions, to small scales in a direct energy cascade \citep{frisch1995turbulence}. In the two-dimensional (2-D) Navier-Stokes equations, both energy and enstrophy are inviscid invariants and this fact constrains the energy transfer to be from small to large scales in an inverse energy cascade \citep{bofetta2012twodimensionalturbulence}. Anisotropic turbulence, such as rotating and stratified turbulence in a finite layer, combines features of the 2-D and 3-D cases. For example, for forced (non-rotating, uniform-density) turbulence in a thin layer, there is a critical value $h_c$ of the parameter $h=H/\ell_{in}$, with layer height $H$ and forcing scale $\ell_{in}$. At $h<h_c$ the flow becomes quasi-2-D and an inverse energy cascade forms \citep{celani2010morethantwo,benavides_alexakis_2017,musacchio2017split,xia2011upscale}.  In this state, part of the injected energy is transferred to larger scales and another part to smaller scales, forming a so-called {\it bidirectional} or {\it split} cascade \citep{alexakis2018cascades}. If the layer has a finite horizontal extent, in the absence of a large-scale damping mechanism, the inverse energy transfer leads to the formation of a condensate, where most of the energy is concentrated at the largest available scale \citep{vankan2019condensates,musacchio2019condensate,vankan2019rare},
\BLUE{a behaviour that has also been confirmed experimentally \citep{xia2009spectrally}.}
Similar transitions from a forward to an inverse cascade and to quasi-2-D motion have also been observed in other systems like magneto-hydrodynamic turbulence \citep{alexakis2011two, seshasayanan2014edge, seshasayanan2016critical} and helically constrained flows \citep{sahoo2015disentangling, sahoo2017discontinuous}, among others (see the articles by \cite{alexakis2018cascades} and \cite{pouquet2019helicity} for recent reviews).

Forced rotating turbulence in fluids of homogeneous density within a layer of finite height displays a similar transition when $\Ro$ is decreased below a threshold $\Ro_c$, giving rise to a split cascade and quasi-2-D flow. The transition to a bidirectional cascade has been studied systematically by \citep{smith1996crossover,deusebio2014dimensional,pestana2019regime}, 
while the transition to a condensate regime was investigated by \citep{alexakis2015rotating, yokoyama2017hysteretic, seshasayanan2018condensates}. \BLUE{Bidirectional energy cascades in rotating turbulent flows have also been measured experimentally \citep{campagne2014direct}.} Recently, \citep{van2020critical} provided evidence that in the limit of simultaneously small $\Ro$ and large $h=H/\ell_{in}$, the transition to a bidirectional cascade occurs at a critical value of the parameter $\lambda=(h\Ro)^{-1}= \lambda_c \approx 0.03$. That study used direct numerical simulations (DNS) of an asymptotically reduced set of equations  derived from the rotating Navier-Stokes equations to achieve extreme parameter regimes that are difficult to reach using a brute-force approach. In the present paper, we extend the results of \citep{van2020critical} to the case of rotating and stably stratified flow.% by introducing density fluctuations on a linear background profile, leading to a constant buoyancy frequency $N$. 

\textcolor{black}{For purely stratified flows, \citep{sozza2015dimensional} provided numerical evidence showing that there is a threshold height $H_c$, below which a split energy cascade appears, with $H_c \propto 1/N$ for $\Fr \ll 1$ and $H\ll \ell_{in}$.} In the case of combined rotating and stratified turbulence, there are numerous investigations reporting the observation of a split energy cascade \citep{smith2002generation,waite2006transition, kurien2008anisotropic,marino2013inverse,marino2014large,rosenberg2015evidence,
marino2015resolving,oks2017inverse,thomas2021forward}. For unstable stratification in the presence of rotation, an inverse cascade has been reported, which leads to the formation of large-scale condensates \citep{favier2014inverse,guervilly2014large,rubio2014upscale,
guervilly2017jets,julien2018impact}. 
%Based on DNS at fixed layer height, \citep{marino2015resolving} argued that the ratio between the inverse energy flux to $\epsilon_-$ and the direct energy flux $\epsilon_+$ is given by $\epsilon_-/\epsilon_+ \propto (\Ro \Fr)^{-1}$. However, they were only able to cover a limited parameter range and the possibility that the transition is critical as a function of $\Ro \Fr$ cannot be excluded \citep{alexakis2018cascades}. 
\BLUE{Despite these numerous studies, little is known for the phase diagram of rotating stratified turbulence. Such a phase diagram is particularly hard to obtain since it implies coverage of the 3-D parameter space ($h,\Ro,\Fr$). Furthermore
if rotation and stratification are misaligned by an angle $\theta$, as is the case for most geophysical applications, a fourth parameter enters the system. It is thus not surprising that  rotating and stratified turbulence is far from understood.  } For instance, it is unknown whether there exists a critical surface separating a bidrectional and forward cascades in this space. To make progress, it is thus worth looking at particular limits.

\BLUE{ Here, we investigate the aligned case $\theta=0$} and focus on the asymptotic regime of deep layers $h\to \infty$ and fast rotation $\Ro\to 0$, with $h \Ro=const.\equiv \lambda^{-1}$, and $\Fr=O(1)$. We rely on an asymptotic expansion, similar to that used in \cite{julien1998new,van2020critical}, which reduces the problem to a 2-D parameter space $(\lambda,\Fr)$. For $\Fr \to \infty$, the problem further simplifies to the purely rotating case studied in \citep{van2020critical}, for which the transition is critical. In the following, we explore the $(\lambda,\Fr)$ parameter space by means of an extensive set of DNS. 

The remainder of this paper is organised as follows. In section \ref{sec:theo_str}, we discuss the theoretical underpinnings of this study, in section \ref{sec:setup} we describe our numerical set-up, in section \ref{sec:num} we present our numerical results. Finally, in section 
\ref{sec:discussion} we discuss our findings and conclude.

%%%%%%%%%%%%%%%%%%%%%%%%%%%%%%%%%%%
\section{Theoretical background}
\label{sec:theo_str}
%%%%%%%%%%%%%%%%%%%%%%%%%%%%%%%%%%%
%In this section we describe the theoretical foundation of our work.
\subsection{From the Boussinesq system to the reduced equations}
\begin{figure}
\centering
\includegraphics[height=7cm,width=3.5cm]{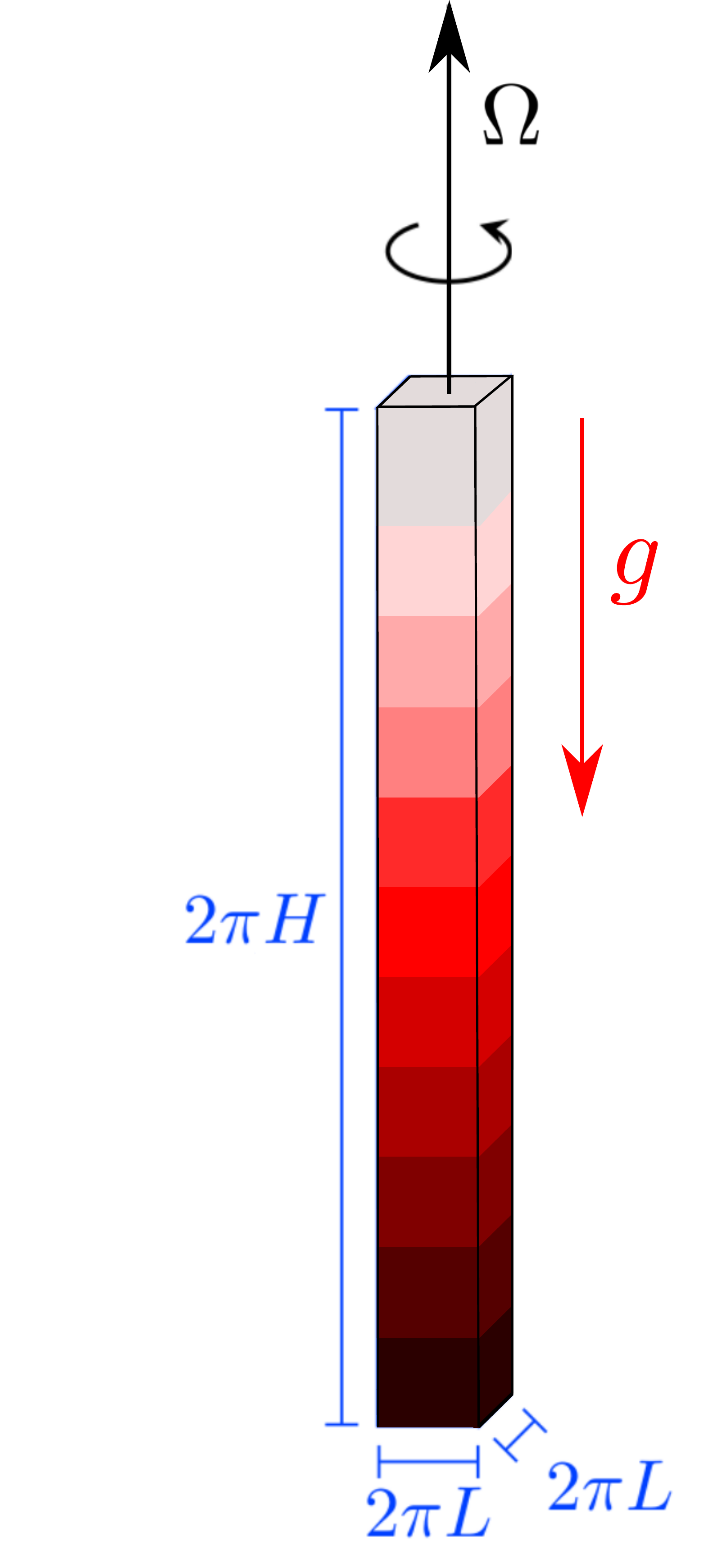}
\caption{The long, rapidly rotating domain with stratification \BLUE{(indicated by grey to black color gradient)}. The black arrow pointing upwards indicates the rotation axis, the red arrow pointing downwards indicates gravity.}
\label{fig:ill}
\end{figure}

The starting point of our investigation is given by the Boussinesq equations in a frame of reference rotating at a constant rate $\boldsymbol{\Omega}=\Omega\hat{e}_\|$, for a linear background density profile $\rho (\mathbf{x},t)= \rho_0 - \alpha (\mathbf{x}\cdot\hat{e}_\|) + \delta \rho (\mathbf{x},t) $, with position $\mathbf{x}$, time $t$, background density $\rho_0=cst.$, stratification strength $\alpha>0$, and $|\rho-\rho_0|\ll \rho_0$. \textcolor{black}{Gravity and stratification are taken to be parallel to the rotation axis.} In their dimensional form, these equations read
\begin{align}
\partial_t \mathbf{u} &\qquad +\mathbf{u}\cdot \nabla\mathbf{u}  &+2\Omega \hat{e}_\|\times \mathbf{u} &=& - \nabla p &+  N \phi \hat{e}_\| &+ \nu \nabla^2 \mathbf{u} &+ \mathbf{f} ,\label{eq:bouss_mom} \\
\partial_t \phi  & \qquad +\mathbf{u}\cdot \nabla \phi& \text{ }&=&  -&N u_\| &+ \kappa \nabla^2 \phi&, \label{eq:bouss_buoy}\\
&&\nabla\cdot \mathbf{u} \text{ } &=&& 0,& &\label{eq:bouss_inc}
\end{align}
with velocity $\mathbf{u}$, pressure (divided by $\rho_0$ \BLUE{and including centrifugal and hydrostatic contributions}) $p$, kinematic viscosity $\nu$, forcing $\mathbf{f}$ (only acting on momentum), buoyancy frequency $N=\sqrt{g\alpha/\rho_0}=cst.$, rescaled density perturbation $\phi(\mathbf{x},t)= N \delta \rho/\alpha$, and diffusivity $\kappa$. The  domain  considered  here  is  the  cuboid  of  dimensions $2\pi L\times 2\pi L\times 2\pi H$, depicted in figure  \ref{fig:ill}, with  periodic boundary  conditions.  For  any  vector $\mathbf{F}$,  we  define  the parallel and perpendicular components as $\mathbf{F}_\| = (\mathbf{F}\cdot \hat{e}_\|) \hat{e}_\|=F_\parallel \hat{e}_\|$, and $\mathbf{F}_\perp =\mathbf{F}-\mathbf{F}_\|$.

%\subsection{The reduced equations of motion}
In the present study, we will explore the regime of simultaneously large $h$ and small $\Ro$ with $\Fr=O(1)$. Brute-force simulations at small $\Ro$ are costly, since very small time steps are required to  resolve fast inertio-gravity waves. Instead,  we  exploit  an  asymptotic  expansion based on the Boussinesq equations, first introduced  in  \citep{julien1998new}, which allows one to investigate the properties of the  transition  to  a  split  cascade in an efficient manner. We consider a stochastic forcing, injecting energy at a constant mean rate into both perpendicular and parallel motions $\langle \mathbf{f}_\perp \cdot \mathbf{u}_\perp\rangle = \langle f_\| u_\|\rangle = \epsilon_{in}/2$, where $\langle \cdot \rangle $ denotes an ensemble average over infinitely many realisations of then noise. The forcing is chosen to be 2-D (independent of the parallel direction), for simplicity, and filtered in Fourier space to act only on a ring of perpendicular wavenumbers centered on $|\mathbf{k}|=k_f=1/\ell_{in}$. A similar 2-D forcing \BLUE{at intermediate lengthscales, smaller than the domain and larger than dissipative scales} has been widely used in previous studies on the transition toward an inverse cascade \citep{deusebio2014dimensional,smith1996crossover,celani2010morethantwo, van2020critical}, \BLUE{and has the attraction of simplicity. In realistic geo- and astrophysical flows, kinetic energy is typically injected in 3-D motions, for instance of convective nature.} %\sout{In the present case, restricting the forcing to the 2-D modes is reasonable because for any forcing with finite correlation time, the injection of energy into the $k_\parallel>0$ modes would be suppressed in the limit  $\Omega\to \infty$, due to the high wave frequencies (to see this, consider a harmonic oscillator forced at a finite time scale as the eigenfrequency diverges). Thus, most of the energy would be injected into the $k_\| =0$ modes.} 
\textcolor{black}{In general, the transition to an inverse cascade can depend on the choice of forcing.} Recent work in thin-layer turbulence by \cite{poujol2020role} suggests that a  3-D forcing,  which  includes  non-zero  parallel wavenumbers, is less efficient at generating an inverse cascade and delays the onset. \BLUE{Furthermore, some recent
results on 2D turbulence with varying $\Rey$ indicated that the nature of the transition can depend on the energy injection mechanism \citep{linkmann2020}. Here, however, we leave investigations on the effect of
the forcing dimensionality for future studies, and focus on the aforementioned 2-D forcing. }

The forcing imposes a length scale $\ell_{in}$, as well as a time scale $\tau_{in}=(\ell_{in}^2/\epsilon_{in})^{1/3}$, and thus a velocity scale $(\epsilon_{in}\ell_{in})^{1/3}$. In terms of these scales, the Rossby number is given by $\Ro = (\tau_{in}\Omega)^{-1}$.
%, and we adopt the $\phi$ scale $\ell_{in}N$. 
The typical scale of parallel variations is $H$, rather than $\ell_{in}$.  Nondimensionalising the equations with these scales, we consider the limit $h=H/\ell_{in}=1/\epsilon$ with $0<\epsilon\ll 1$ and $\Ro=O(\epsilon)$, such that $\lambda=(h\Ro)^{-1}=O(1)$ is independent of $\epsilon$. A multiple-scale expansion \citep{sprague2006numerical}, or a heuristic derivation analogous to that presented in \citep{van2020critical}, can  be  used  to obtain a set of asymptotically reduced equations for the parallel components of velocity $u_\parallel$ and vorticity $\omega_\| =  (\nabla \times \mathbf{u})\cdot \hat{e}_\|$, whose dimensionless form reads
\begin{align}
\partial_t u_\| &+ \mathbf{u}_\perp \cdot \nabla_\perp u_\|   &=& -2 \lambda \partial_\| \nabla_\perp^{-2} \omega_\|  &- \frac{1}{\Fr} \phi + \frac{1}{\Rey} \nabla^2_\perp u_\|& + f_\|, \label{eq:mom_z_red}\\
\partial_t \omega_\|& + \mathbf{u}_\perp \cdot \nabla_\perp \omega_\|  &=& +2\lambda \partial_\| u_\|& + \frac{1}{\Rey}\nabla_\perp^2 \omega_\|& + f_\omega, \label{eq:vort_z_red} \\
\partial_t \phi& + \mathbf{u}_\perp \cdot \nabla_\perp \phi  &=& &\frac{1}{\Fr}  u_\| + \frac{1}{\Pe} \nabla_\perp^2 \phi&, \label{eq:buoy_red}
\end{align}
where $\partial_\| = \hat{e}_\|\cdot \nabla$, $\nabla_\perp = \nabla - \hat{e}_\| \partial_\|$, and the nondimensional parameters $\Fr = (\epsilon_{in} /\ell_{in}^2)^{1/3}/N$, $\lambda = (h \Ro)^{-1} = \ell_{in}^{5/3} \Omega / (\epsilon_{in}^{1/3} H)$, $\Rey = (\epsilon_{in}\ell_{in}^4)^{1/3}/\nu$, $\Pe=(\epsilon_{in}\ell_{in}^4)^{1/3}/\kappa$, \BLUE{and $f_\omega = \hat{e}_\| \cdot (\nabla \times \mathbf{f})$}.  The perpendicular velocity $\mathbf{u}_\perp$ is divergence-free to leading order, $\nabla_\perp \cdot \mathbf{u}_\perp = 0$, which permits us to write it in terms of a stream function $\psi$, such that $\mathbf{u}_\perp = \hat{e}_\| \times \nabla \psi$, and $\omega_\|= \nabla_\perp^2 \psi$. These nondimensional equations are valid in the rescaled domain $2\pi\Lambda\times2\pi\Lambda\times 2\pi$, with $\Lambda=L/\ell_{in}$. Importantly, in eqs. (\ref{eq:mom_z_red}) and (\ref{eq:vort_z_red}), all the information about $H, \Omega$ is contained in the single parameter $\lambda$.

\subsection{Conservation laws}
%For infinite $\Rey$ and $\Pe$ 
In the inviscid and non-diffusive case ($\nu=\kappa=0$), the system conserves the total energy $\mathcal{E} = \frac{1}{2} \int (\mathbf{u}^2 + \phi^2)d^3x$. In addition, the \textit{potential vorticity}
\begin{equation}
q = 2\lambda \partial_z \phi - \omega_\|/\Fr + (\partial_y u_\|)(\partial_x\phi) - (\partial_x u_\|)(\partial_y \phi) \label{eq:potvort}
\end{equation}
 (in Cartesian coordinates, with the parallel direction being $z$) is conserved along each fluid parcel trajectory. \textcolor{black}{Eq. (\ref{eq:potvort}) is a simplified, Boussinesq version of Ertel's full potential vorticity \citep{ertel1942neuer} (the full form applies to compressible flow).} The material conservation of $q$ implies that $\mathcal{C}_n = \int q^n d^3x$ is conserved for all $n$, where the special case $n=2$ is known as \textit{potential enstrophy}. In 2-D turbulence, energy and enstrophy are both quadratic functionals of the stream function, with enstrophy containing higher spatial derivatives. The simultaneous conservation of the two quantities constrains the energy cascade to be to larger scales, and the enstrophy to smaller scales. By contrast, $\mathcal{C}_2$ is not directly related to the kinetic or potential energy, and does not imply a straightforward constraint for cascade directions, except in a special case, which shall be discussed later.
 %However, since $q^2$ is not quadratic in the fields, it does not provide a simple constraint on the energy cascades in general, unlike vorticity in 2-D turbulence. This is no longer true for $\Fr \ll 1$, in which case $q$ is dominated by the term $\omega_\|/\Fr$, such that $q^2$ becomes quadratic. Then an inverse cascade of kinetic energy in the perpendicular components of velocity is possible (BUT THE ENERGY IS NOT PURELY KINETIC... DOES THIS ARGUMENT WORK?). 
 
Eqs. (\ref{eq:mom_z_red}),  (\ref{eq:vort_z_red}) and (\ref{eq:buoy_red}) are closely related to well-known models in geophysical fluid dynamics. Since the leading-order perpendicular velocity is in geostrophic balance, and only the perpendicular velocity appears in the advection terms, the model resembles the classical quasi-geostrophic equations valid in thin layers \citep{pedlosky2013geophysical}. Indeed (\ref{eq:mom_z_red}-\ref{eq:buoy_red}) have been referred to as \textit{generalised quasi-geostrophic} equations \citep{julien2006generalized}. Variants of the reduced equations have been applied in a variety of contexts, such as rotating turbulence \citep{nazarenko2011critical}, rapidly rotating convection \citep{sprague2006numerical,grooms2010model,
julien2012heat,julien2012statistical,
rubio2014upscale,maffei2021inverse},
as well as dynamos driven by rapidly rotating convection \citep{calkins2015multiscale}.

%%%%%%%%%%%%%%%%%%%%%%%%%%%%%%%%%%%%%%%%%%%%%%%%%%%%%%%%
\subsection{Inertio-gravity waves and slow modes}
\label{sec:igws}
A fundamental property of rotating and stratified flows is that they support inertio-gravity waves. In the full Boussinesq equations (\ref{eq:bouss_mom}-\ref{eq:bouss_inc}), the dispersion relation of these waves reads 
\begin{equation}
\sigma^2(\mathbf{k}) = \frac{4\Omega^2 k_\|^2 + N^2 k_\perp^2}{k^2}, \label{eq:disp_igw}
\end{equation}
where $\sigma$ is the wave frequeny, $\Omega$ is the rotation rate, $N$ is the buoyancy frequency, $\mathbf{k}$ is the wave vector, $k_\|$ is the component of the wave vector along the rotation axis, $k_\perp$ the component perpendicular to the rotation axis, and $k^2 = k_\|^2 + k_\perp^2$. In the framework of the reduced equations of motion (\ref{eq:mom_z_red}-\ref{eq:buoy_red}), this simplifies, in nondimensional form, to
\begin{equation}
\sigma^2(\mathbf{k}) = 4 \lambda^2 \frac{k_\parallel^2}{k_\perp^2} + \frac{1}{\Fr^2} \label{eq:disp_igw_red},
\end{equation}
where $\sigma$ and the wavenumber components are nondimensional. At large $\Omega$, (\ref{eq:disp_igw}) implies high wave frequencies, requiring a small time step to be resolved numerically. In the reduced equations, all parameters are of order one, which makes numerical simuation more efficient \BLUE{by filtering the fast inertio-gravity waves}.

The full set of linear modes of rotating stratified flow has been studied in great detail \citep{leith1980nonlinear, bartello1995geostrophic,sukhatme2008vortical,herbert2014restricted}. Here we just summarise some relevant results. Formally, linearising (\ref{eq:mom_z_red}-\ref{eq:buoy_red}), one obtains an equation of the form $\dot{ \mathbf{Z}}(\mathbf{k}) = \mathbf{L}(\mathbf{k}) \mathbf{Z}(\mathbf{k})$, with $\mathbf{Z}(\mathbf{k})=(k_\perp\hat{\psi}(\mathbf{k}),\hat{u}_\parallel(\mathbf{k}),\hat{\phi}(\mathbf{k}))$ with hats denoting Fourier transforms, and a $3\times3$ matrix $\mathbf{L}$. The eigenvalues of $\mathbf{L}$ are $+\sigma(\mathbf{k)},-\sigma(\mathbf{k}),0$, with $\sigma(\mathbf{k})>0$ given by eq. (\ref{eq:disp_igw_red}). Thus, in addition to waves with frequencies $\pm \sigma$, one also finds linear eigenmodes with zero frequency at every wavenumber. The corresponding normalised eigenvector is  
\begin{equation}
\mathbf{Z}_0(\mathbf{k}) = \frac{1}{\sigma(\mathbf{k}) k_\perp}\left(-i k_\perp \Fr^{-1},\text{ }0, \text{ }2\lambda k_\| \right),
\end{equation} 
which notably has a vanishing $\hat{u}_\parallel$ component. These \textit{slow} modes with zero frequency span the so-called \textit{slow manifold}. The normalised eigenvectors of $\mathbf{L}$ with eigenvalues $\pm \sigma(\mathbf{k})$ are
\begin{equation}
\mathbf{Z}_\pm(\mathbf{k}) = \frac{1}{\sqrt{2} \sigma(\mathbf{k}) k_\perp}\left(2 \lambda k_\|, \text{ }\pm\sigma(\mathbf{k}) k_\perp,\text{ }-i k_\perp \Fr^{-1}\right).
\end{equation}
which has a nonvanishing $\hat{u}_\parallel$ component. \textcolor{black}{We highlight that the wave modes have zero potential vorticity at the linear level. The slow modes are thus the vortical modes of the flow.}

 In order for wave modes to interact efficiently with the slow modes, the inverse wave frequency of the slowest waves must be comparable to the eddy turnover time scale of the turbulent 2-D flow $\tau_{in}$. In the purely rotating case ($\Fr\to \infty$), this argument was successfully used by \cite{van2020critical} to predict the dependence of the energy cascades on $\lambda$: forward cascade at $\lambda<\lambda_c\approx 0.03$ and inverse cascade at $\lambda > \lambda_c$. For the rotating and stratified case, two cases can be anticipated based on (\ref{eq:disp_igw_red}).
\subsection{Weak stratification: the passive-scalar limit}
 At weak stratification ($\Fr>1$), the system is likely to be close to the purely rotating case, such that a transition should occur when $\lambda>\lambda_c(\Fr)$. While we do not predict the dependence of $\lambda_c(\Fr)$, one expects that $\lambda_c= (H_c \Ro_c)^{-1}$ increases with stratification. This is because as the weak stratification is increased (while remaining weak), kinetic and potential energy become more strongly coupled, and more kinetic energy will be converted to potential energy, which behaves approximately like a passive scalar at weak stratification. \textcolor{black}{For passive scalars, it is well known that scalar variance (potential energy) cascades forward (to small scales) \citep{warhaft2000passive,falkovich2001part,celani2004active}. Therefore stratification will counteract the inverse cascade. Thus it appears reasonable that faster rotation, i.e. higher $\lambda$, should be required at weak stratification for generating an inverse energy flux. A similar effect has been observed in thin-layer turbulence, where a decrease of the critical height has been observed with increased stratification \citep{sozza2015dimensional}. }
 
%%%%%%%%%%%%%%%%%%%%%%%%%%%%%%%%%%%%%%%%%%%%%%%%%%%%%%%%
\subsection{Strong stratification: the hydrostatic limit}
For strong stratification ($\Fr \ll 1$) and large $\lambda$, the dominant balance in (\ref{eq:mom_z_red}) is given by
\begin{equation}
2\lambda \partial_\| \nabla_\perp^{-2} \omega_\|  = - \phi/\Fr. \label{eq:hydrost_balance}
\end{equation} 
%\textcolor{black}{Modes in the slow manifold correspond to balanced motion in the sense that they satisfy (\ref{eq:hydrost_balance}) at the linear level, even though nonlinearity can disrupt the balance.} Formally, the hydrostatic limit can be studied by letting
  Eq. (\ref{eq:hydrost_balance}) is a form of hydrostatic balance, which is common in geophysical flows \citep{vallis2017atmospheric}. To see this, one can identify the stream function of the perpendicular flow as $\nabla_\perp^{-2} \omega_\|=\psi$, which follows from $\mathbf{u}_\perp = \hat{e}_\parallel \times \nabla \psi$. Comparing the latter relation to geostrophic balance, between Coriolis force and perpendicular pressure gradient, one further deduces that $\psi$ is proportional to the pressure. Hence eq. (\ref{eq:hydrost_balance}) is a balance between the vertical (parallel) pressure gradient and gravity, i.e. hydrostatic balance.

\BLUE{Note that for $\lambda \gg Fr^{-1}$ the dynamic hydrostatic balance just corresponds to a two-dimensionalization of the flow. This is because hydrostatic balance implies small $\partial_\| \psi$ in this limit. %, and the omega equation constrains $u_\|$ to be small, such that that the flow is approximately two-dimensional with two components.
However, when $Fr$ is of order one or smaller the flow is not necessarily 2-D. For small or $O(1)$ values of $\lambda$, the dynamic hydrostatic balance limit is expected to hold when the wave frequency is much larger than typical eddy turn over time, i.e. $\Fr\ll 1$. We highlight the fact that the combination $\lambda \Fr \propto \Fr/\Ro \propto \Omega/N$, which has been indentified as a control parameter in previous studies \citep{smith2002generation,marino2015resolving}, appears naturally here in eq. (\ref{eq:hydrost_balance}).}

We note that modes in the slow manifold defined in section \ref{sec:igws} correspond to balanced motion in the sense that they satisfy (\ref{eq:hydrost_balance}) at the linear level. At the nonlinear level, even if the flow starts at hydrostatic balance,  
its nonlinear evolution can disrupt it. However, in the limit of high wave frequencies one can expect that the inertio-gravity waves will decouple from the slow manifold, which will therefore evolve independently, always satisfying eq. (\ref{eq:hydrost_balance}). Such a limit can be formally captured by letting $\lambda \to \lambda/\epsilon$, $\Fr^{-1}\to Fr^{-1}/\epsilon$, $u_\| \to \epsilon u_\|$, with $\epsilon \ll 1$, while $\omega_\|, \phi \to \omega_\|,\phi$. 
\BLUE{This is the stratified quasi-geostrophic (QG) limit \citep{charney1971geostrophic}: the potential vorticity defined in eq. (\ref{eq:potvort}) simplifies to give
\begin{equation}
q = -\Fr^{-1} \left[ (2\lambda \Fr)^2 \partial_\|^2 + \nabla_\perp^2  \right] \psi + O(\epsilon) \equiv - \Fr^{-1} \overline{\nabla}^2 \psi + O(\epsilon),
\end{equation}
where we identified the rescaled Laplace operator $\overline{\nabla}^2 \equiv (2\lambda \Fr)^2 \partial_\|^2 + \nabla_\perp^2  $, which contains $2\lambda\Fr=\Omega \ell_{in}/(N H)=Bu^{-1/2}$, where $Bu$ is the Burger number \citep{cushman2011introduction}. The QG potential vorticity remains conserved along particle trajectories in the inviscid unforced case, i.e.
\begin{equation}
    (\partial_t + \mathbf{u}_\perp \cdot \nabla_\perp )q = 0 \text{ } (+\text{forcing and dissipation})
\end{equation}
The theory of QG dynamics is commonly discussed in much detail in textbooks on geophysical fluid dynamics, such as  \cite{salmon1998lectures} and chapter 5 of \cite{vallis2017atmospheric}. A particular advantage of this formulation is the inversion principle: only a single scalar variable $q$ needs to be advected, which gives $\psi$ by inverting the elliptic operator $\overline{\nabla}^2$, and thus $\mathbf{u}_\perp$ by geostrophic balance, $\phi$ from hydrostatic balance, and $u_\|$ can be found by combining these relations in the form of an omega equation \citep{hoskins1978new,hoskins2003omega}. In the QG limit, at leading order
\begin{equation}
\mathcal{C}_2 = \Fr^{-2} \int  \left(\overline{\nabla}^2 \psi\right)^2 d^3x + O(\epsilon)
\end{equation}
while the total energy becomes, at leading order,
\begin{equation}
\mathcal{E} = \frac{1}{2} \int  \left(\overline{\nabla}\psi\right)^2 + O(\epsilon^2).
\end{equation}
and it is well known since the early contributions of \cite{charney1971geostrophic,rhines1979geostrophic,salmon1980baroclinic} that turbulent QG flow produces an inverse energy cascade as a result. This has also been confirmed by numerical simulations \citep{hua1986numerical,mcwilliams1989statistical,vallgren2010charney}. Since $\psi$ and $\phi$ are directly linked by the hydrostatic balance, both kinetic and potential energy cascade inversely.}

\if 
We note that taking the time derivative of (\ref{eq:hydrost_balance}) gives the intriguing diagnostic (as opposed to prognostic) relation
\begin{equation}
\overline{\nabla}^2 u_\| =  S, \label{eq:omega_equation}
\end{equation}
with %the elliptic operator $\mathcal{L} \equiv (\Fr^{-2} \nabla_\perp^2 + 4 \lambda^2 \partial_\|^2)$ and 
the source term
\begin{equation}
S = \Fr^2\left\lbrace\Fr^{-1} \nabla_\perp^2 [(\mathbf{u}_\perp \cdot \nabla_\perp)\phi] + 2 \lambda  \partial_\| [(\mathbf{u}_\perp\cdot \nabla_\perp) \omega_\|]\right\rbrace.
\end{equation}
\textcolor{black}{Equation (\ref{eq:omega_equation}) is a variant of the \textit{omega equation}, which is well known (for an arbitrary stable background density profile) in meteorology \citep{hoskins1978new,hoskins2003omega}. In the meteorological literature, pressure is often taken as the vertical coordinate in an ideal-gas atmosphere. In this coordinate system, the vertical velocity is conventionally denoted by $\omega$ by meteorologists, hence the name of the equation.} The omega equation has been used extensively for diagnosing vertical velocities, which are highly relevant for weather phenomena, from observed vorticity and temperature fields. %\sout{The role of $u_\|$ in eqns. (\ref{eq:mom_z_red}), (\ref{eq:vort_z_red}) for hydrostatically balanced situations is similar to the relation between pressure and the velocity field in incompressible turbulence, in that $u_\|$ no longer an independent variable.}
\fi

%%%%%%%%%%%%%%%%%%%%%%%%%%%%%%%%%%%
\section{Numerical set-up and methodology}
\label{sec:setup}
%%%%%%%%%%%%%%%%%%%%%%%%%%%%%%%%%%%
In this section, we describe the numerical set-up used in the present study. The partial differential equations that we solve numerically in a domain $2\pi \Lambda \times 2\pi \Lambda\times 2\pi$  are given by (\ref{eq:mom_z_red}), (\ref{eq:vort_z_red}) and (\ref{eq:buoy_red}) with modified dissipative terms
\begin{align}
&\partial_t u_\| &+ \mathbf{u}_\perp \cdot \nabla_\perp u_\| &+  2 \lambda \partial_\| \nabla_\perp^{-2} \omega_\| &=& - \frac{\phi}{\Fr} &&- \frac{(-\nabla_\perp^2)^{n} u_\| }{\Rey_\perp}  - \frac{(-\partial_\|^2)^{m} u_\|}{\Rey_\|}  + f_\|, \label{eq:mom_z_red_num}\\
&\partial_t \omega_\| &+ \mathbf{u}_\perp \cdot \nabla \omega_\| &- 2\lambda \partial_\| u_\| &=& &&-\frac{(-\nabla_\perp^2)^n \omega_\|}{\Rey_\perp} -\frac{(-\partial_\|^2)^m \omega_\|}{\Rey_\|}+ f_\omega, \label{eq:vort_z_red_num} \\
&\partial_t \phi &+ \mathbf{u}_\perp \cdot \nabla \phi & &=&  +\frac{u_\|}{\Fr}&  &- \frac{(- \nabla_\perp^2)^{n_\phi} \phi}{\Pe_\perp}- \frac{ (-\partial_\|^2)^{m_\phi}\phi}{\Pe_\|}.\hspace{1.5cm} \label{eq:buoy_red_num}
\end{align}
Note that there is no large-scale friction term, such that an inverse cascade can develop unhindered and accumulate energy the scale of the box. Moreover, the density perturbation field $\phi$ is not forced directly in our simulations. As in \citep{van2020critical}, the parallel dissipation terms, which do not appear in (\ref{eq:mom_z_red}),  (\ref{eq:vort_z_red}) and (\ref{eq:buoy_red}), are added for numerical reasons, suppressing the formation of exceedingly large parallel wavenumbers. We choose the hyperviscosity exponents $n=m=n_\phi=m_\phi=4$ for all simulations. 

Equations (\ref{eq:mom_z_red_num}-\ref{eq:buoy_red_num}) are controlled by seven nondimensional parameters. In addition to $\Lambda$, $\lambda$ and $\Fr$, which are defined identically to  eq. (\ref{eq:mom_z_red}-\ref{eq:buoy_red}), there are two Reynolds numbers and two Peclet numbers associated with perpendicular and parallel diffusion terms, respectively:
\begin{equation}
\Rey_\perp = \frac{\epsilon_{in}^{1/3} \ell_{in}^{2n-2/3}}{ \nu_n}, \hspace{0.1cm} \Rey_\| = \frac{\epsilon_{in}^{1/3} \ell_{in}^{2m-2/3}}{\mu_m},\hspace{0.1cm} \Pe_\perp=\frac{\epsilon_{in}^{1/3} \ell_{in}^{2n_\phi-2/3}}{\kappa_n}, \hspace{0.1cm} \Pe_\| = \frac{\epsilon_{in}^{1/3} \ell_{in}^{2m_\phi-2/3}} {\kappa_{m_\phi}}
\end{equation} 
with hyperviscosities $\nu_n,\nu_m$ and hyperdiffusivities $ \kappa_{n_\phi},\kappa_{m_\phi}$. 

We solve equations (\ref{eq:mom_z_red_num}-\ref{eq:buoy_red_num}) in the triply periodic domain using a pseudo-spectral code based on the Geophysical High-order Suite for Turbulence, including $2/3$-aliasing (see \cite{mininni2011hybrid}). A total of 71 runs were performed at a resolution of $512^3$ with $\Lambda=32$, of which 63 runs at $\Rey_\perp = \Rey_\|= \Pe_\perp = \Pe_\perp\| = 9200$, for different values of $\Fr$ and $\lambda$, and an additional 8 runs at $\Rey_\|=\Pe_\|=4600$ halved, with $\Rey_\perp, \Pe_\perp$ unchanged, to verify that our results do not depend on the parallel dissipation terms added for numerical reasons. \BLUE{For completeness, one run was also performed at $512^2\times 1024$ and $\Rey_\perp = \Rey_\|= \Pe_\perp = \Pe_\perp\| = 9200$ to verify well-resolvedness, and another at $512^2\times 1024$, and $\Rey_\|=\Pe_\|=18400$ with $\Rey_\perp, \Pe_\perp$ unchanged, verifying that the results are independent of $\Rey_\|,\Pe_\|$.}\\

In order to characterise the energy cascades, we measure several quantities in every run, which are defined below, with hats indicating Fourier transforms. The 2-D kinetic energy spectrum is defined as 
\begin{equation}
E_{kin}(k_\perp,k_\parallel) = \frac{1}{2} \sum_{k_\perp-\frac{1}{2} \leq p_\perp <k_\perp + \frac{1}{2}} \left(\frac{|\hat{\omega}_\parallel(\mathbf{p}_\perp,k_\|)|^2}{k_\perp^2}  + |\hat{u}_\parallel^2(\mathbf{p}_\perp,k_\|)|^2 \right), \label{eq:Ekin_spec_2D}
\end{equation}
and the 2-D potential energy spectrum as
\begin{equation}
E_{pot}(k_\perp,k_\parallel) = \frac{1}{2} \sum_{k_\perp-\frac{1}{2} \leq p_\perp <k_\perp + \frac{1}{2}} |\hat{\phi}(\mathbf{p}_\perp, k_\|)|^2, \label{eq:Epot_spec_2D}
\end{equation}
where hats denote Fourier transforms. The one-dimensional (1-D) energy spectrum is obtained by summing the 2-D spectra over $k_\parallel$,
\begin{align}
E_{kin}(k_\perp) =& \sum_{k_\|} E_{kin}(k_\perp,k_\|) \equiv E_{kin}^\perp(k_\perp) + E_{kin}^\|(k_\perp), \label{eq:Ekin_spec_1D} \\
E_{pot}(k_\perp) =& \sum_{k_\|} E_{pot}(k_\perp,k_\|), \label{eq:Epot_spec_1D}
\end{align}
where $E_{kin}^\perp$ contains all terms involving $\hat{\omega}_\|$ and $E_{kin}^\|$ contains all terms involving $\hat{u}_\|$. In addition, we define the total energy spectrum $E_{tot}= E_{kin}+E_{pot}$.

The 2-D dissipation spectra are defined as
\begin{align}
D_{kin}(k_\perp,k_\|) = &\sum_{k_\perp-\frac{1}{2} \leq p_\perp <k_\perp + \frac{1}{2}} (\nu_n p_\perp^{2n} + \nu_m k_\|^{2m})  \left(\frac{|\hat{\omega}_\parallel(\mathbf{p}_\perp,k_\|)|^2}{k_\perp^2}  + |\hat{u}_\parallel^2(\mathbf{p}_\perp,k_\|)|^2 \right), \label{eq:diss_spec_kin} \\
D_{pot}(k_\perp,k_\|) =& \sum_{k_\perp-\frac{1}{2} \leq p_\perp <k_\perp + \frac{1}{2}}  (\kappa_{n_\phi} p_\perp^{2n_\phi} + \kappa_{m_\phi} k_\|^{2m_\phi}) |\hat{\phi}(\mathbf{p}_\perp, k_\|)|^2, \label{eq:diss_spec_pot}
\end{align}
giving the total dissipation spectrum $D_{tot}=D_{kin}+D_{pot}$. Finally, the spectral energy fluxes in the perpendicular direction through a cylinder of radius $k_\perp$ in Fourier space are defined as
\begin{align}
\Pi_{kin}^\perp (k_\perp) =& \langle (u_\perp)^<_{k_\perp} \cdot [(\mathbf{u}_\perp \cdot \nabla_\perp) \mathbf{u}_\perp] \rangle, \\
\Pi_{kin}^\| (k_\perp) =& \langle (u_\|)^<_{k_\perp} [(\mathbf{u}_\perp\cdot \nabla_\perp) u_\|]\rangle, \\
\Pi_{pot} (k_\perp ) =& \langle \phi^<_{k_\perp} [(\mathbf{u}_\perp \cdot \nabla_\perp) \phi]\rangle,
\end{align}
with the total energy flux defined as $\Pi_{tot} \equiv \Pi_{kin}^\perp+\Pi_{kin}^\|+\Pi_{pot}$, where for any field $A$,
\begin{equation}
A^<_{k_\perp}(\mathbf{x}) \equiv \sum_{\stackrel{\mathbf{p}}{p_\perp<k_\perp}} \hat{A}(\mathbf{p})\exp(i\mathbf{p}\cdot \mathbf{x}).
\end{equation}

Every run is initialised at a random small-energy configuration, and continued until  
\begin{itemize}
\item[1.] an inverse energy flux is observed, with kinetic energy piling up at the large scales,
\item[2.] or a purely forward cascade is observed and the system has reached steady state.
\end{itemize}

%%%%%%%%%%%%%%%%%%%%%%%%%%%%%%%%%%%
\section{Simulation results}
\label{sec:num}
%%%%%%%%%%%%%%%%%%%%%%%%%%%%%%%%%%%
%In this section we present the results of our simulations.
%%%%%%%%%%%%%%%%%%%%%%%%%%%%%%%%%%%%%%%
\subsection{Overview of parameter space}
%%%%%%%%%%%%%%%%%%%%%%%%%%%%%%%%%%%%%%%
\begin{figure}
\centering
\includegraphics[width=0.49\textwidth]{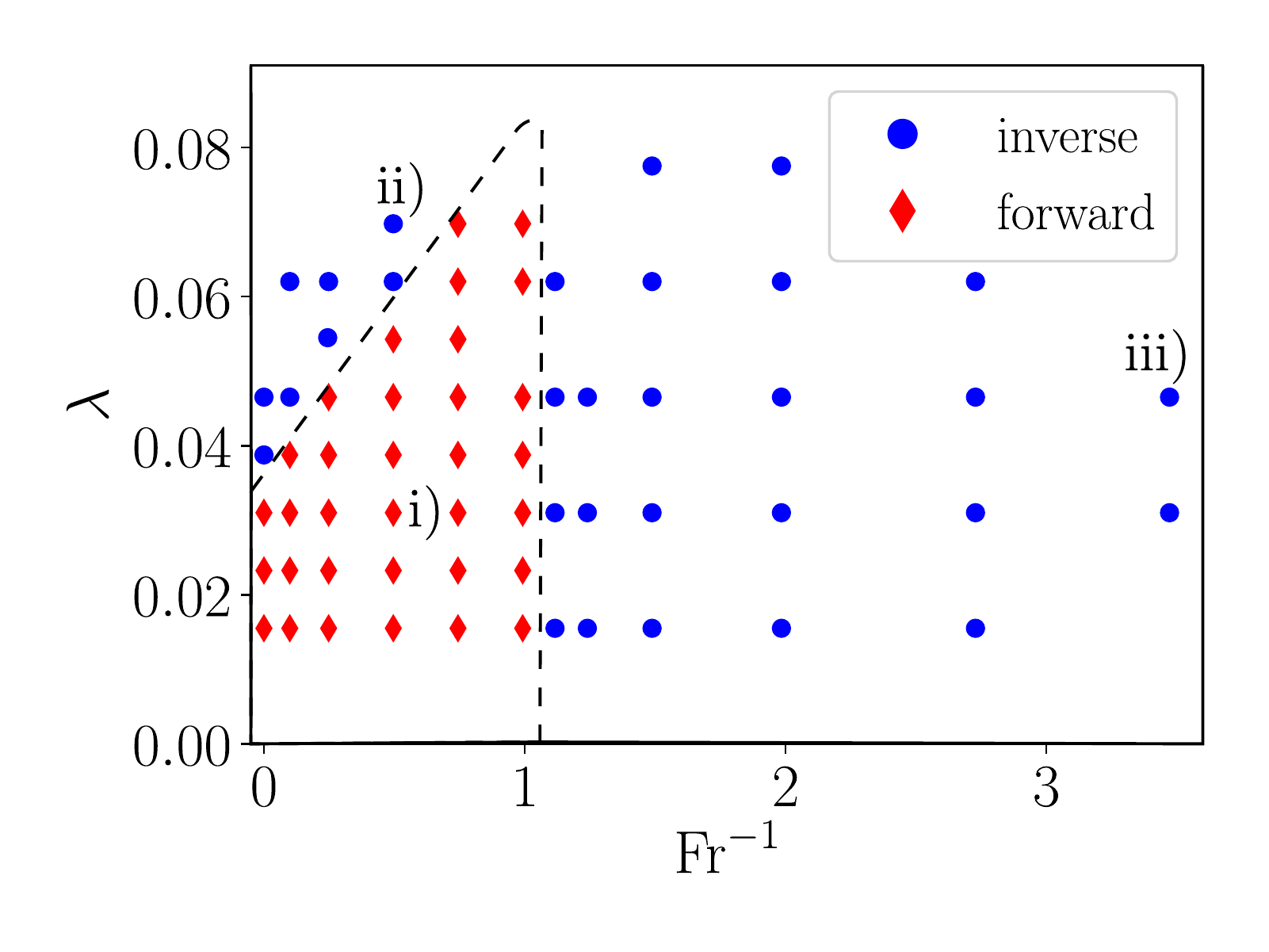}
\includegraphics[width=0.49\textwidth]{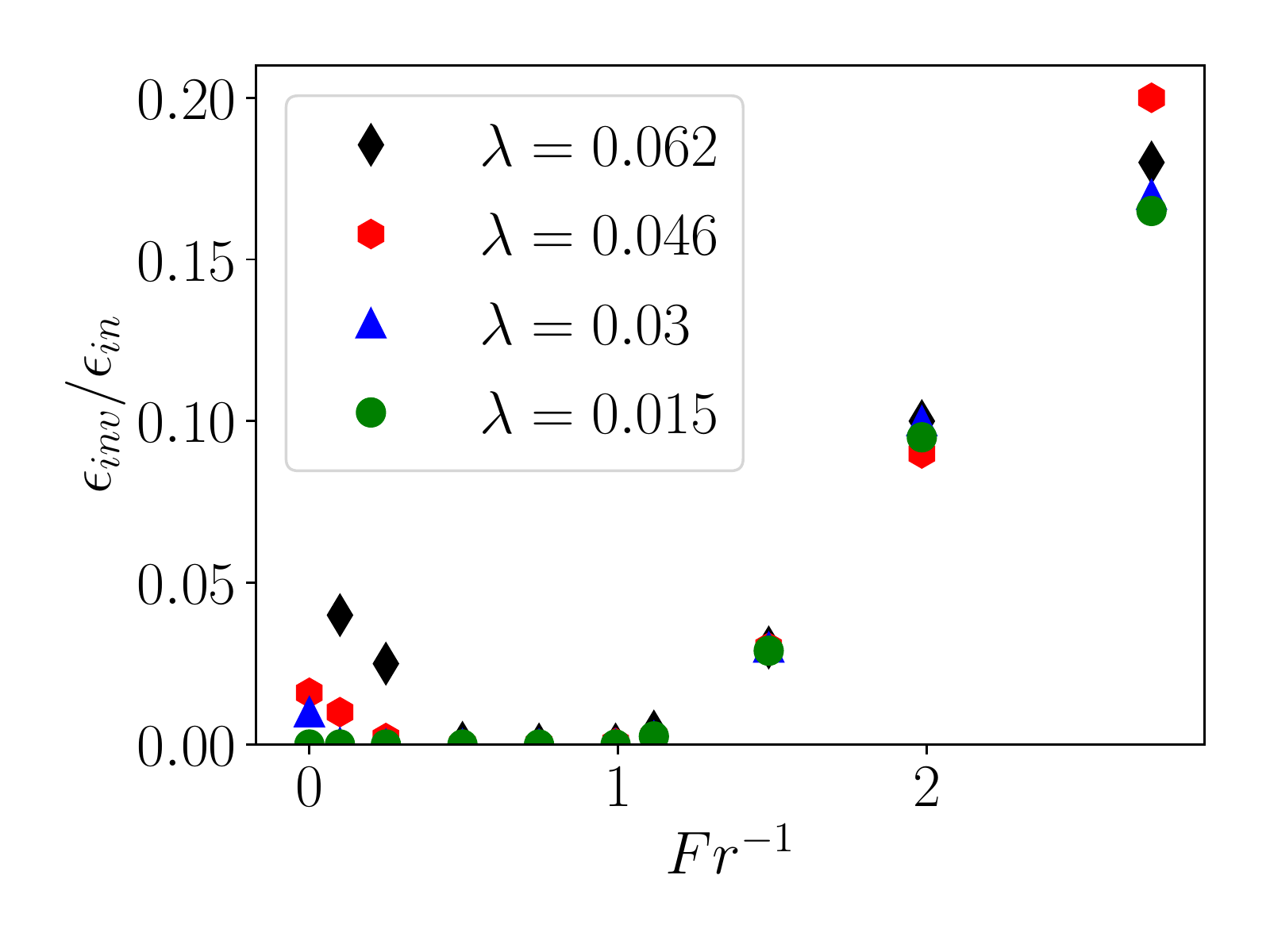}
{ \caption{ \raggedright Left: Regime diagram showing the direction of the kinetic energy cascade for various values of the parameters $(\lambda,\Fr^{-1})$. A tentative boundary between forward and split cascading states is shown by the dashed line. The labels i), ii) and iii) indicate the three states to be examined in more detail below. \BLUE{Right: fraction of injected energy cascading inversely versus $\Fr^{-1}$ at four different values of $\lambda$.} }\label{fig:overview}
}
\end{figure}
First we provide an overview of the runs. \BLUE{ The left panel of figure \ref{fig:overview} shows} a regime diagram indicating for which values of $\lambda$ and $\Fr$ an inverse cascade in kinetic energy was observed. Two regions can be discerned: a finite region (red diamonds) near the origin in terms of $(\lambda,\Fr^{-1})$, where an only-forward-cascading state is observed, and a surrounding region (blue circles) at larger $\lambda$ (faster rotation / shallower box) and larger $\Fr^{-1}$ (strong stratification), where an inverse energy cascade arises. The boundary between the two is tentatively shown by the dashed lines. 
\BLUE{The right panel of the same figure( \ref{fig:overview}) shows the rate energy energy cascades to the large scales $\epsilon_{inv}$ normalized by the energy injection rate $\epsilon_{in}$ as a function of $Fr^{-1}$ for four different values of $\lambda$. Simulations with $\epsilon_{inv}>0.01 \epsilon_{in}$ were titled as inverse cascading in the left panel. Note that the transition from one state to the other appears to be sharp, although further investigations would be required to determine the behavior close to the onset of the inverse cascade.  }

The boundary between the two regions is consistent with our \textcolor{black}{expectations} from section \ref{sec:theo_str}: first, for $\Fr>1$ (weaker stratification), there is a (roughly linear) increase in $\lambda_c$, i.e. the critical value of $(\Ro h)^{-1}$, with $\Fr^{-1}$. While we do not offer a theoretical prediction for the linear scaling, an identical scaling $h_c\propto 1/N$ has been suggested for strongly stratified turbulence in a thin layer \citep{sozza2015dimensional}. Second, when $\Fr$ is lowered beyond $\Fr\approx 1$, \textcolor{black}{the system enters the hydrostatic regime, and} a direct energy cascade turns into an inverse cascade.  \BLUE{ We note that the quasi-geostrophic limit strictly applies for large $\lambda$, while the boundary in figure \ref{fig:overview} appears independent of $\lambda$ and the inverse cascade persist even for 
small values of $\lambda$. The inverse cascade predicted for the quasi-geostrophic limit appears thus to extend beyond its range of validity.
This behavior is possibly related to the isolation of the slow modes when the inertio-gravity waves become very fast, which occurs for $\Fr^{-1}\gg1$ independently of the value of $\lambda$, based on eq. \ref{eq:disp_igw_red}.}

%While the threshold $\Fr$ appears to be independent of $\lambda$ in our simulations, we cannot exclude the possibility that this situation is different at lower $\lambda$ than we could reach.
%%%%%%%%%%%%%%%%%%%%%%%%%%%%%%%%%%%%%%%
\subsection{Spectra}
\begin{figure}
\includegraphics[width=0.325\textwidth]{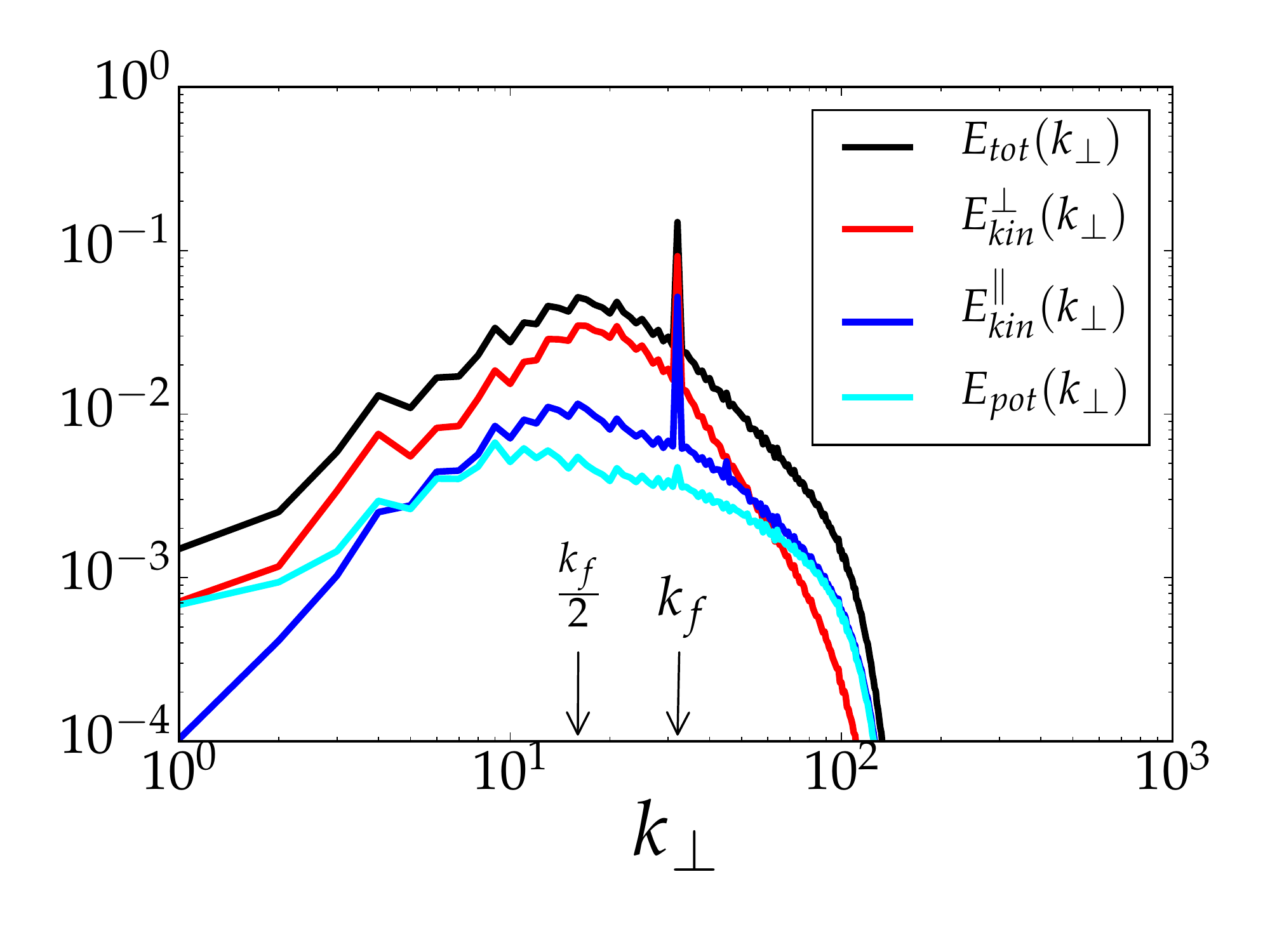}
\includegraphics[width=0.325\textwidth]{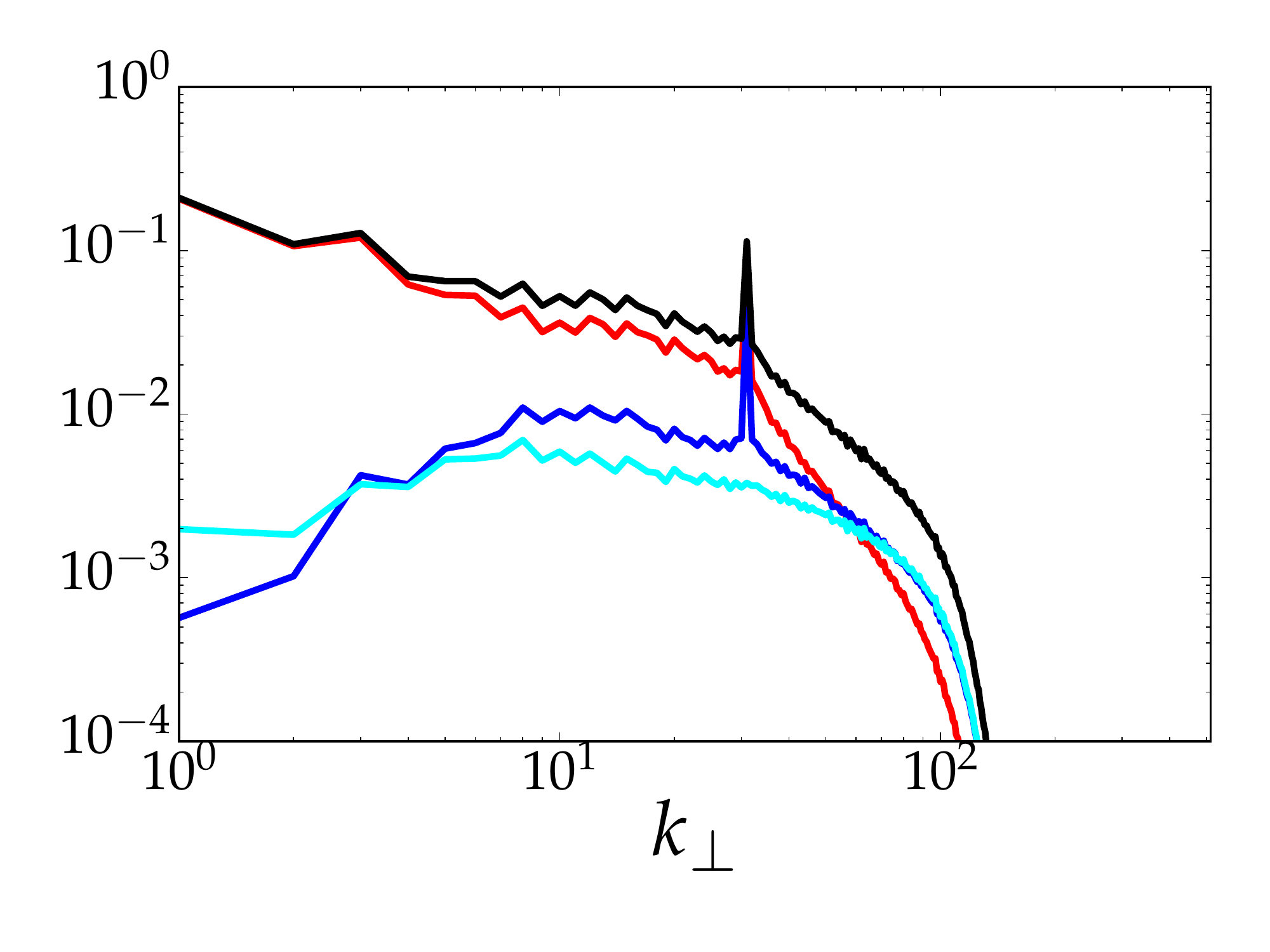}
\includegraphics[width=0.325\textwidth]{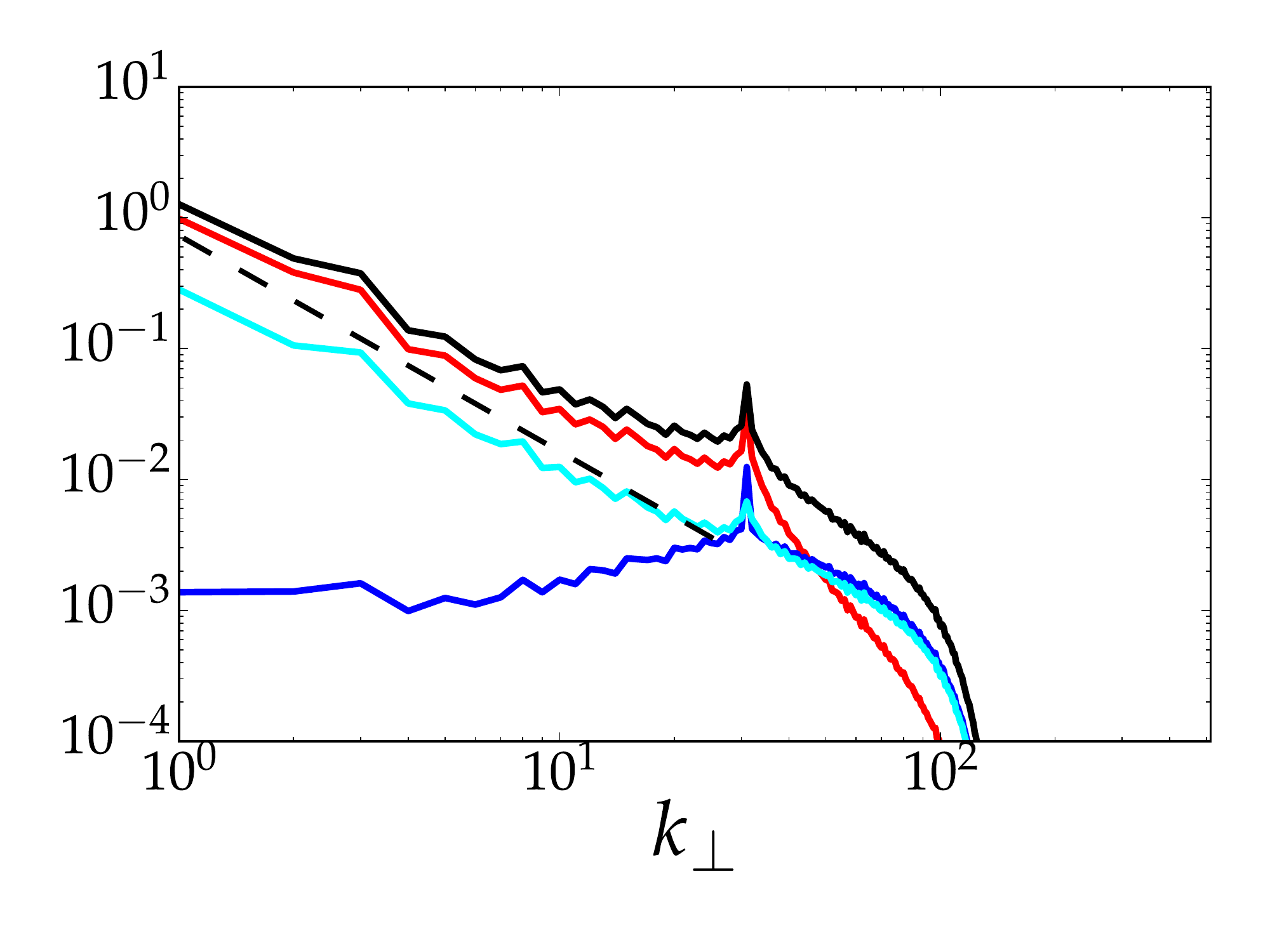}
\caption{Double logarithmic plots of the contributions to the 1-D energy spectra according to equations (\ref{eq:Ekin_spec_1D}) and (\ref{eq:Epot_spec_1D}). Left: case i), center: case ii), right: case iii). The legend on the left applies to all panels. \BLUE{The dashed line shows a $-5/3$ power-law for reference.} }
\label{fig:1D_spectra}
\end{figure}

%%%%%%%%%%%%%%%%%%%%%%%%%%%%%%%%%%%%%%%
In the following, we illustrate three different representative cases highlighted in figure \ref{fig:overview},
\begin{itemize}
\item[i)] $\lambda = 0.03$ $\Fr^{-1}=0.5$ (no inverse cascade),
\item[ii)] $\lambda = 0.07$, $\Fr^{-1}= 0.5$ (weak stratification, inverse cascade), 
\item[iii)] $\lambda =0.045$, $\Fr^{-1}= 3.5$ (strong stratification, inverse cascade).
\end{itemize}
\BLUE{The results shown below are from the simulations at $512^3$. The simulation results at higher resolution and different Reynolds and P\'eclet numbers showed no qualitative differences.}
The 1-D energy spectra are shown in figure \ref{fig:1D_spectra}. For case i), in the forward-cascading regime, there is a spectral maximum in both perpendicular and parallel kinetic energy at $k_\perp\approx k_f/2$. A similar phenomenon is reported in \citep{van2020critical} for the purely rotating case, where an instability mechanism was suggested as the cause for this secondary maximum. The potential energy spectrum is peaked at yet larger scales $k_\perp < k_f/2$. While we do not offer a theoretical explanation for the local spectral maxima at scales larger than the forcing scale, the similarity with the phenomenology of the rotating case suggests that a related instability mechanism may be at play. The potential energy spectrum is comparable to the parallel kinetic energy spectrum, except at the largest scales, where it is comparable to the perpendicular kinetic energy, and at the forcing scale, where it is  smaller, since potential energy is not directly forced. For case ii), where an inverse cascade is present at weak stratification, the perpendicular kinetic energy spectrum shows a maximum at  the largest scale $k_\perp=1$, where it dominates the total energy. The parallel kinetic energy and and the potential energy, by contrast, do not show a maximum at the largest scales. Finally, in case iii), where an inverse cascade is present at strong stratification, both the perpendicular kinetic energy spectrum and the potential energy spectrum shows maximum at $k_\perp=1$, with a clear power-law range at $k_\perp<k_f$. The shape of the potential energy spectrum is strikingly similar to the perpendicular kinetic energy, only differing by constant factor of around $0.3$ over two decades in $k_\perp$. One also observes a peak at the forcing scale, although the potential energy is not directly forced. These observations indicate that the density field and the parallel vorticity are non-trivially related to each other for all scales but the very smallest. As discussed in section \ref{sec:theo_str}, this can occur as the consequence of hydrostatic balance. This will be examined in section \ref{sec:vis}. In case iii), the parallel kinetic energy does not show a secondary maximum. \BLUE{We stress that in cases ii) and iii), the results shown are from the transient state where the inverse cascade continues to develop, by contrast with case i) where a stationary state is reached.}

\begin{figure}
\includegraphics[width=0.325\textwidth]{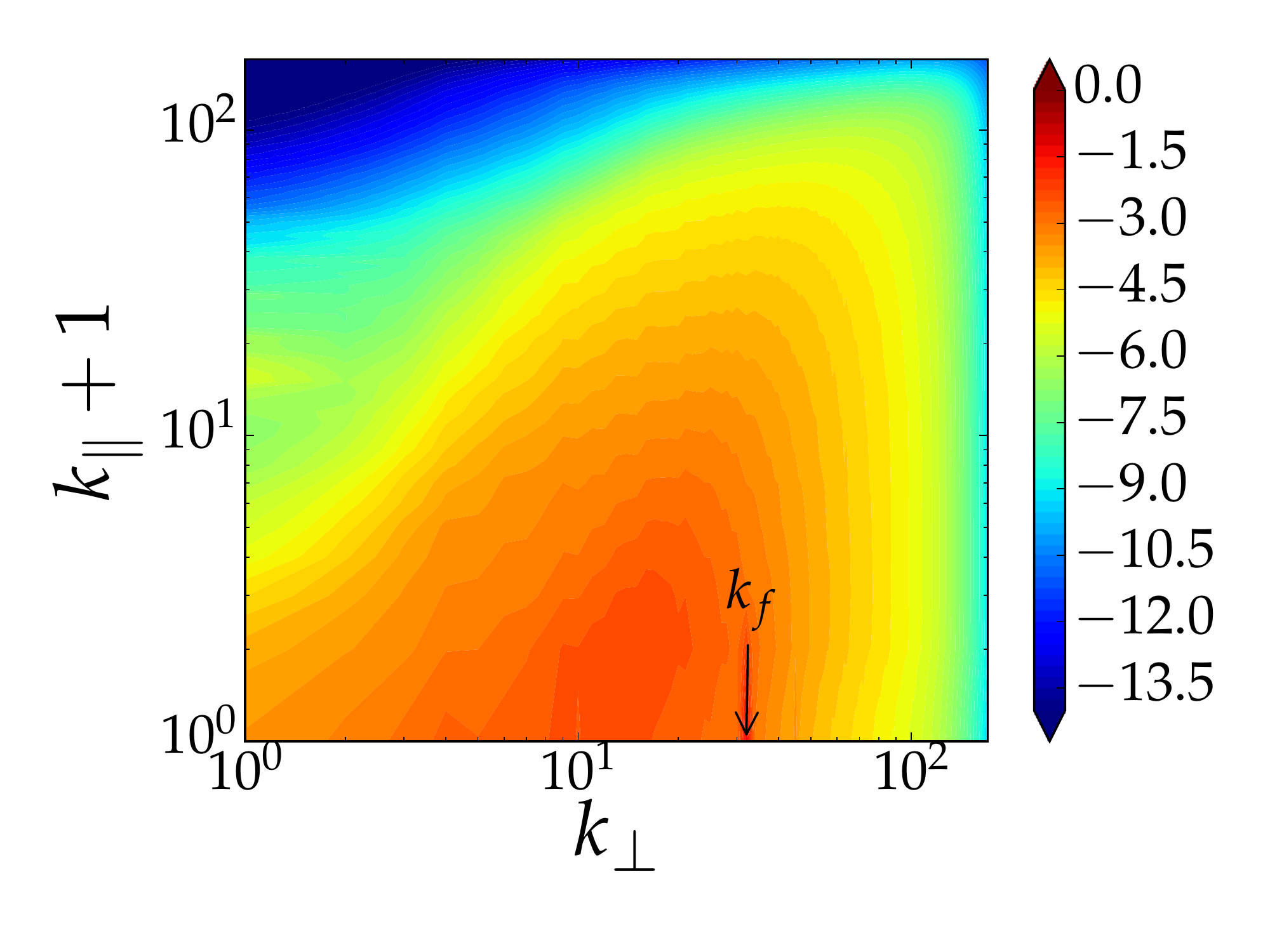}
\includegraphics[width=0.325\textwidth]{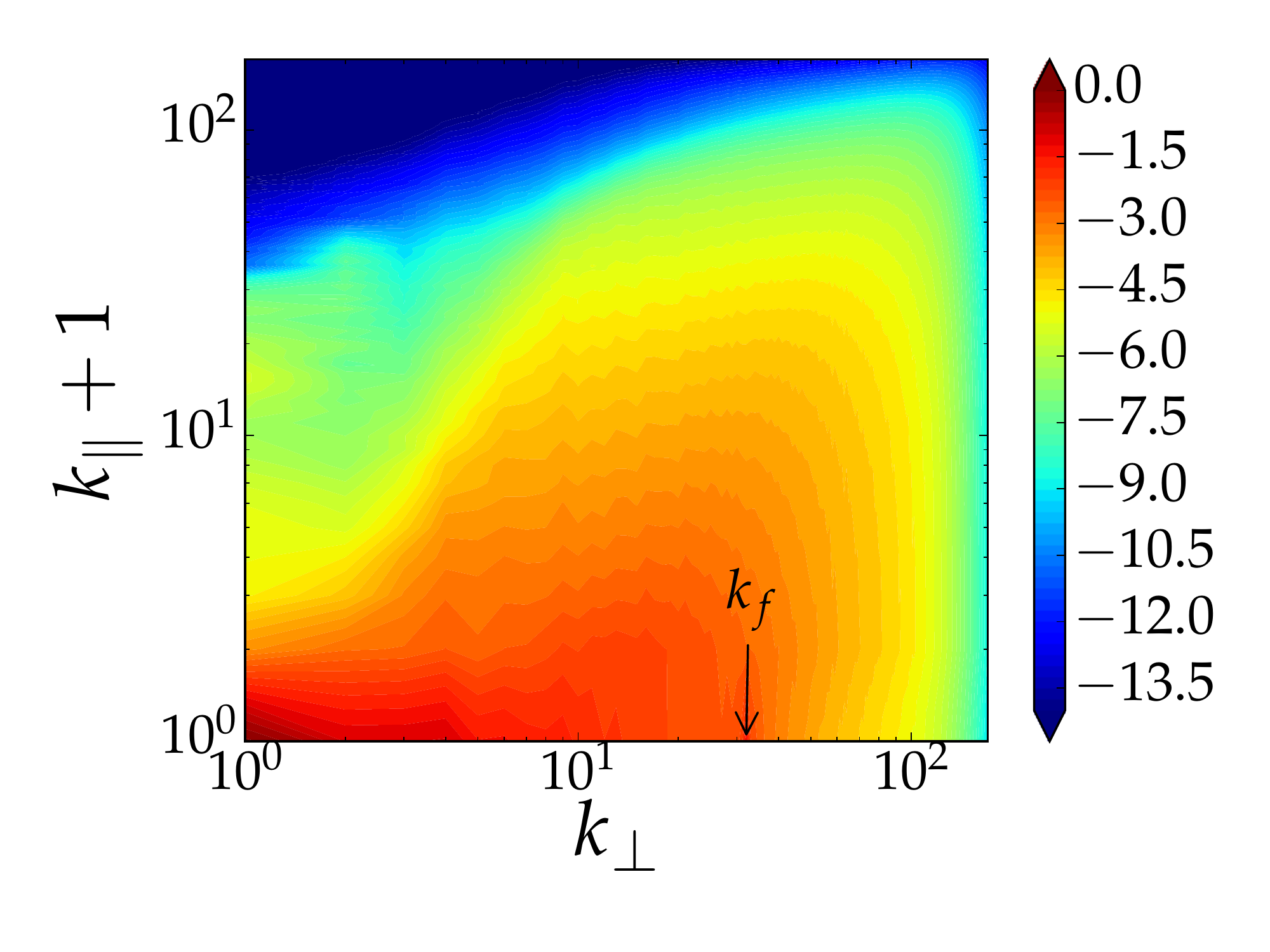}
\includegraphics[width=0.325\textwidth]{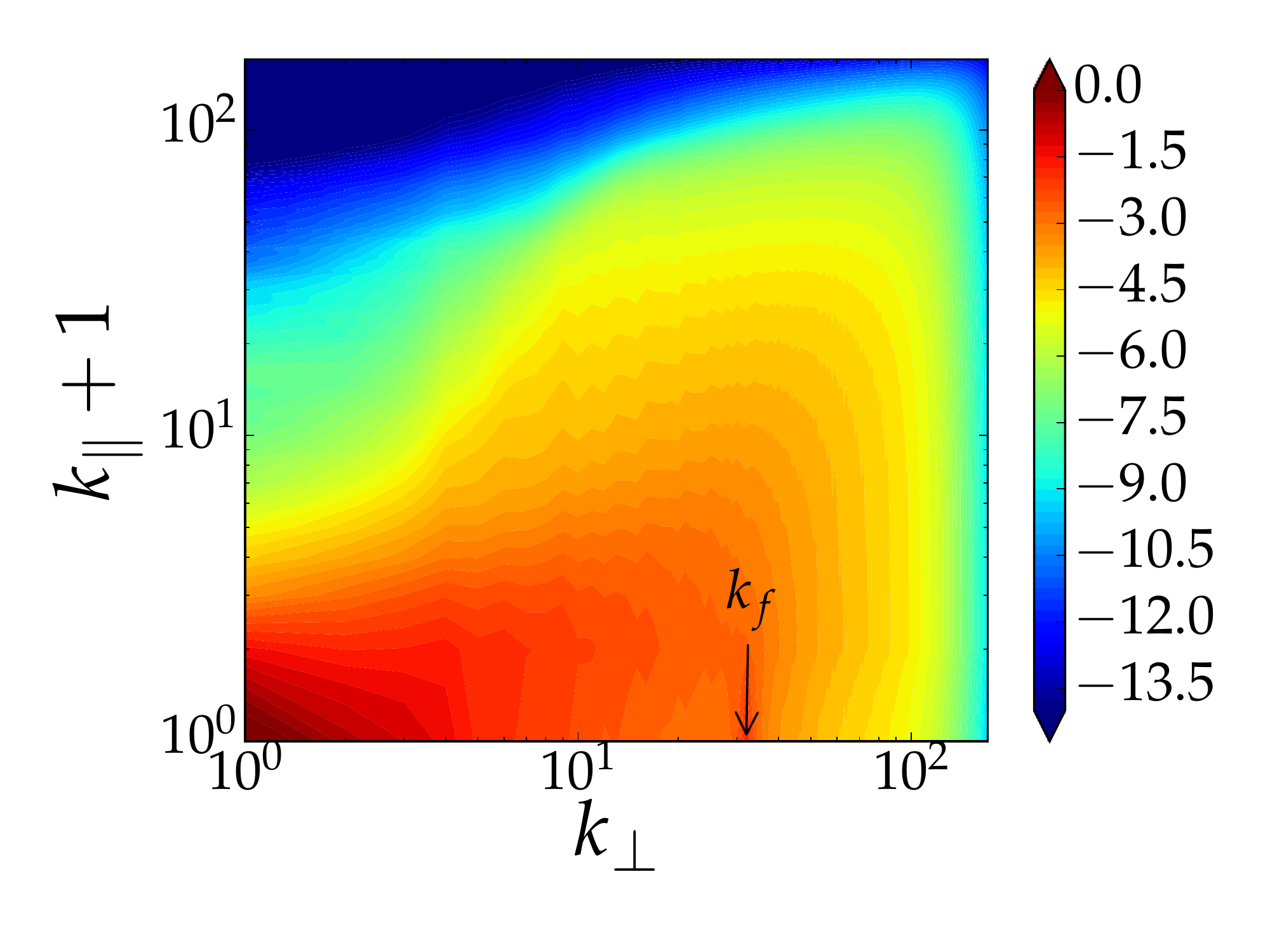}
\caption{Filled contour plots of the 2-D kinetic energy spectrum $E_{kin}(k_\perp,k_\|)$ defined in equation (\ref{eq:Ekin_spec_2D}), as a function of $k_\perp$, $k_\|$ for cases i) to iii) (left to right).}
\label{fig:2Dspec_Ekin}
\end{figure}

\begin{figure}
\includegraphics[width=0.325\textwidth]{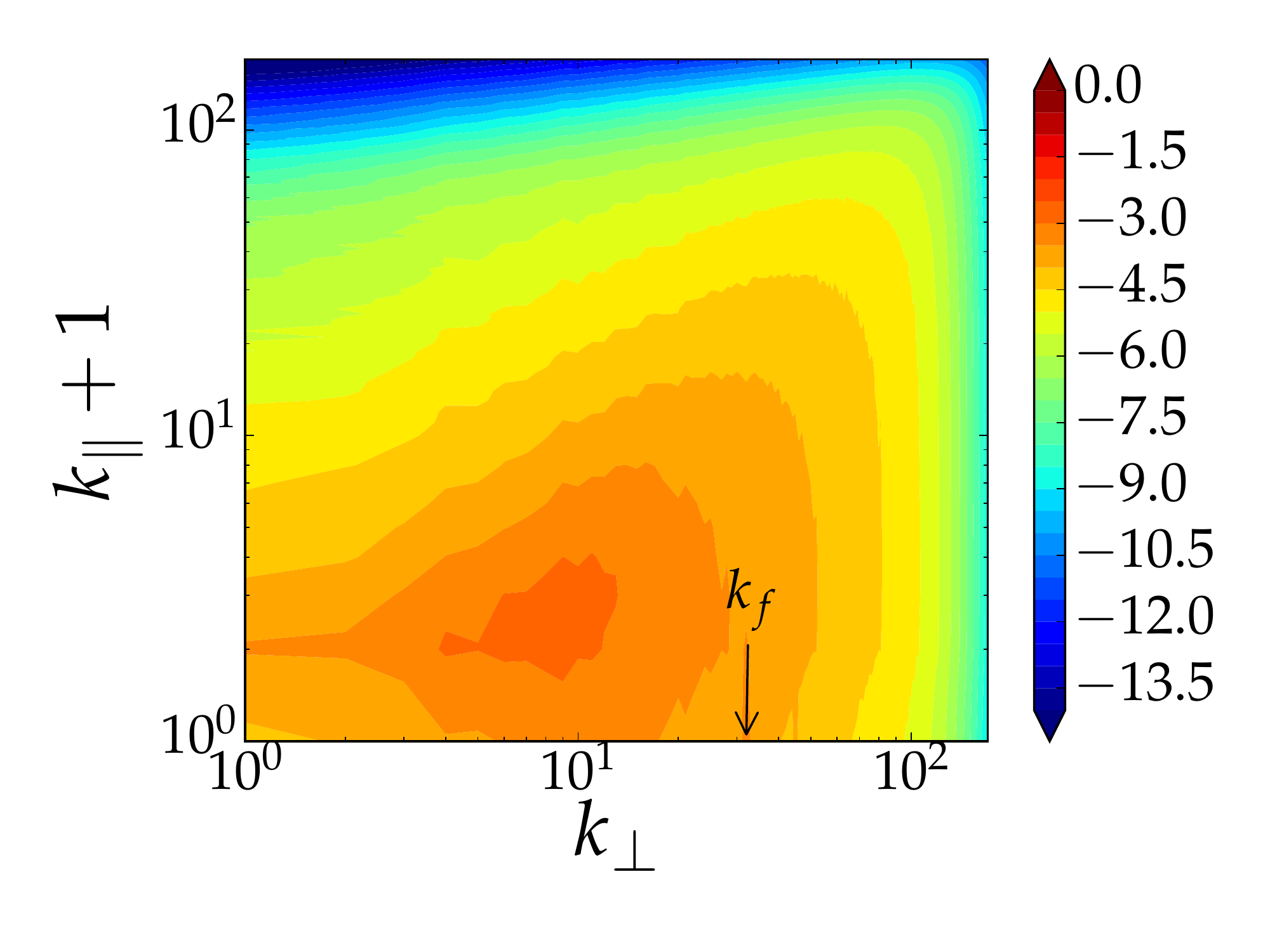}
\includegraphics[width=0.325\textwidth]{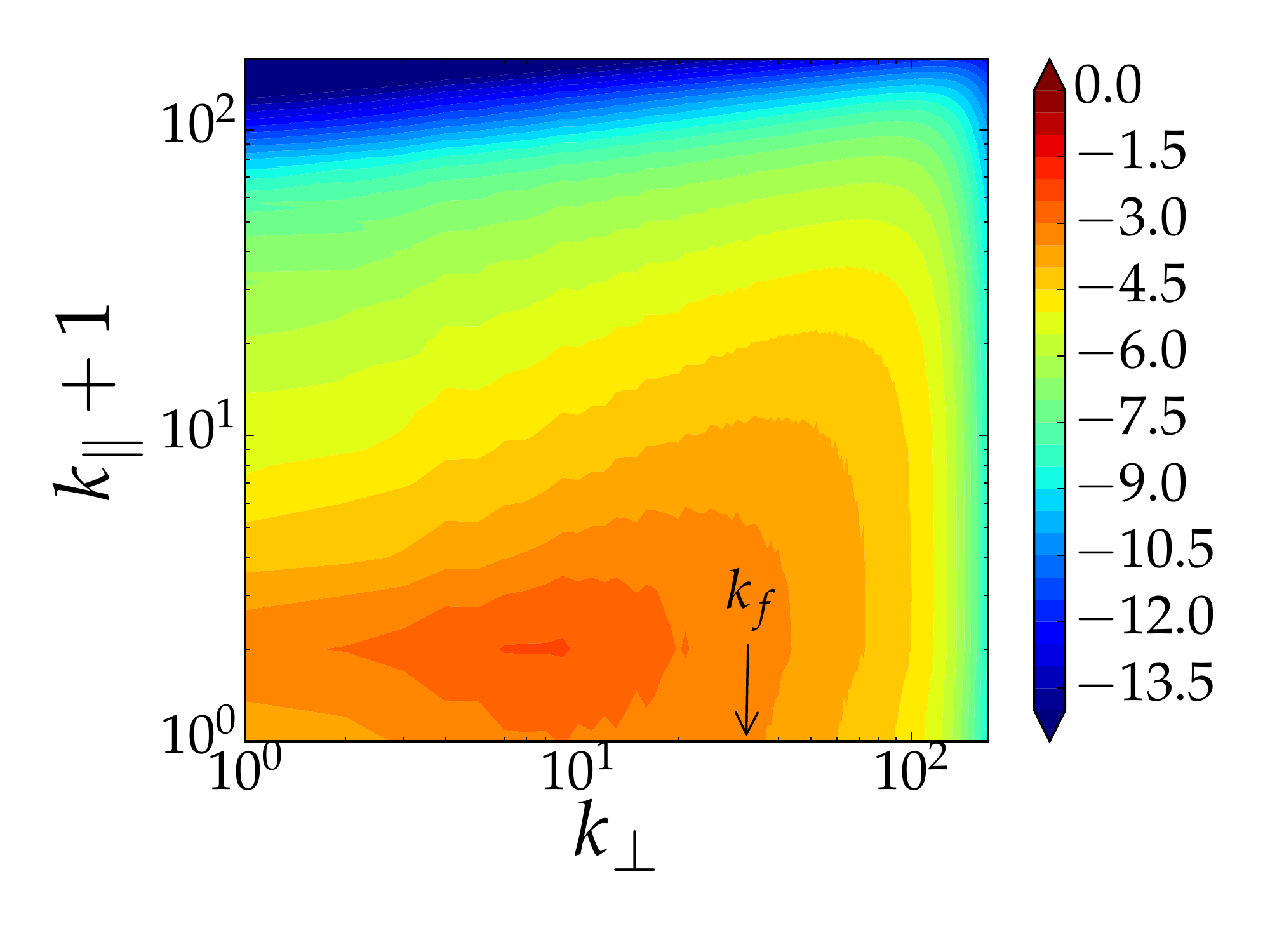}
\includegraphics[width=0.325\textwidth]{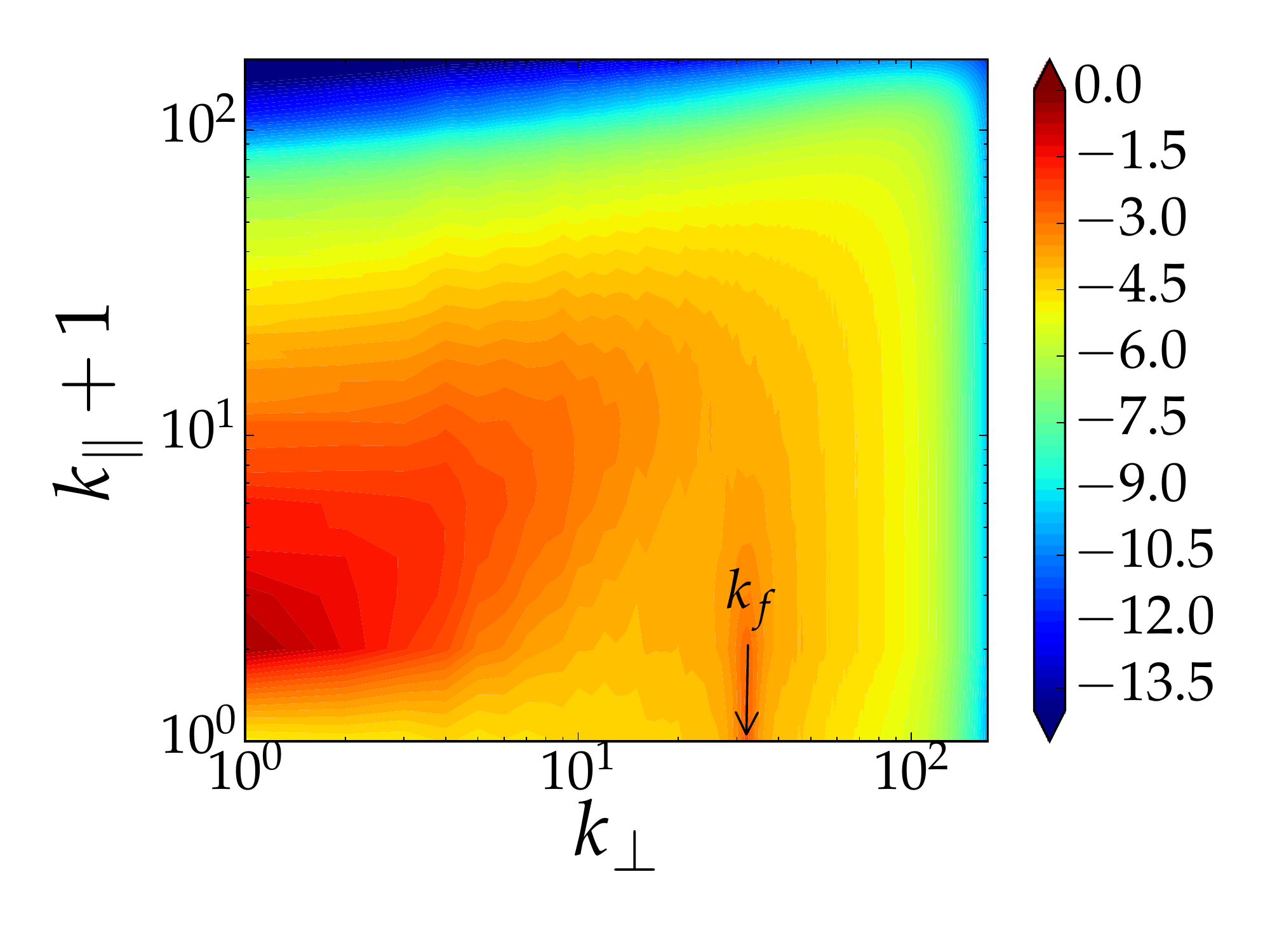}
\caption{Filled contour plots of 2-D potential energy spectrum $E_{pot}(k_\perp,k_\|)$ defined in (\ref{eq:Epot_spec_2D}) versus $k_\perp$, $k_\|$ for cases i) to iii) (left to right). }
\label{fig:2Dspec_Epot}
\end{figure}

The 2-D kinetic energy spectra (sum of perpendicular and parallel contributions) are shown in figure \ref{fig:2Dspec_Ekin}. In case i), the spectral maximum at $k_\perp\approx k_f/2$ is seen to extend to $k_\parallel >0$. In cases ii) and iii), the spectral maximum at $k_\perp=1$ is seen to stem primarily from contributions at $k_\parallel=0$. The 2-D potential energy spectrum is shown figure \ref{fig:2Dspec_Epot}. For cases i) and ii), there is a maximum at intermediate $k_\perp$, with $k_\| =1$. By contrast, for case iii) there is a clear build-up of potential energy at $k_\perp=1$, and maximum at $k_\|=1$ (and some contributions from $k_\|=2$). In case iii), there is only little potential energy at $k_\|=0$, even though the kinetic energy spectrum peaks at $k_\|=0$, which is compatible with hydrostatic balance (\ref{eq:hydrost_balance}). 
%%%%%%%%%%%%%%%%%%%%%%%%%%%%%%%%%%%%%%%
\subsection{Energy fluxes}
\begin{figure}
\includegraphics[width=0.325\textwidth]{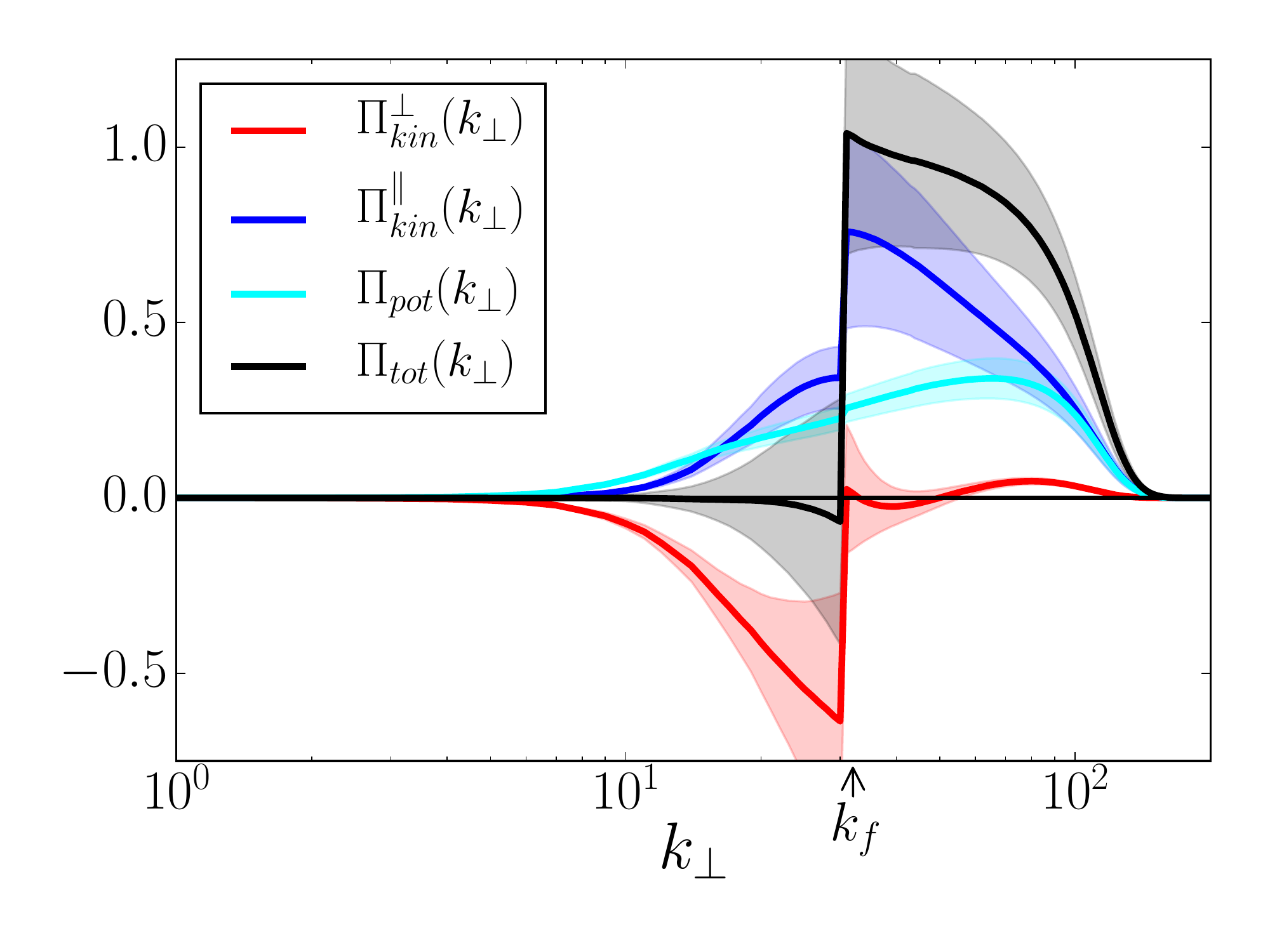}
\includegraphics[width=0.325\textwidth]{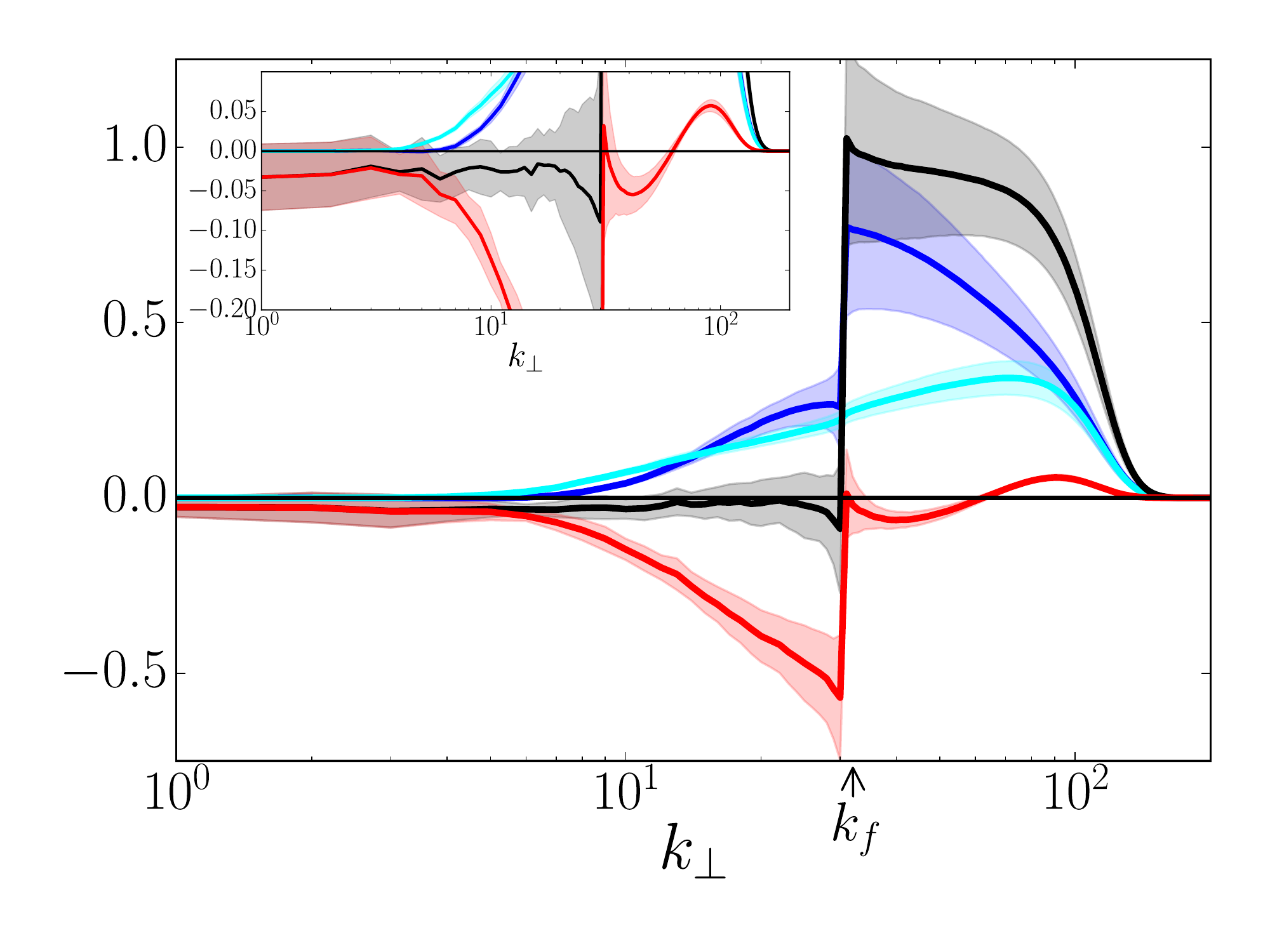}
\includegraphics[width=0.325\textwidth]{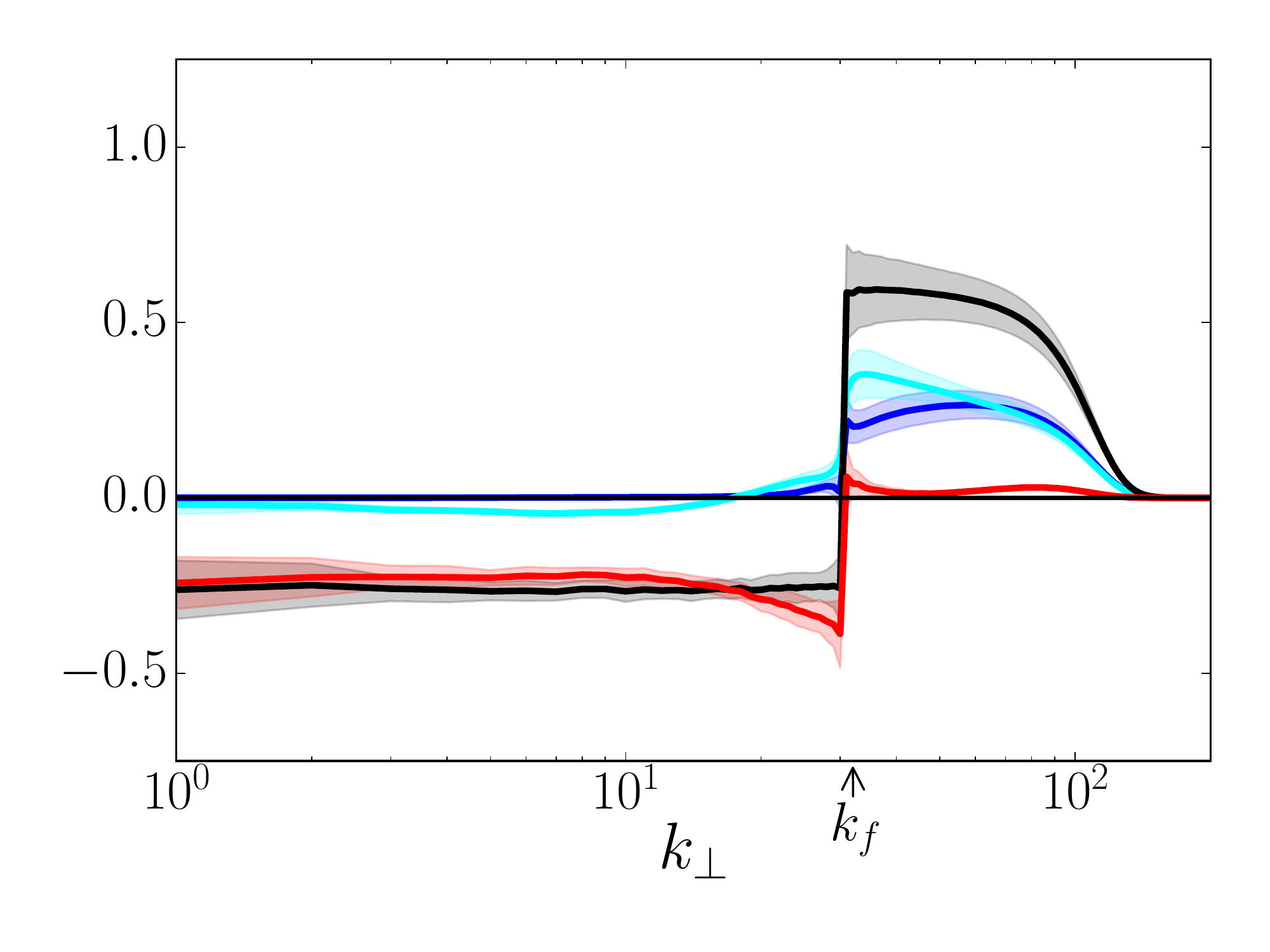}
\caption{Average energy fluxes (non-dimensionalised by the injection rate $\epsilon$) for cases i (left), ii (center) and iii) right. The shaded area around curves shows one standard deviation of fluctuations about the average. In the central panel, the inset shows a zoom on negative-flux range. }
\label{fig:fluxes}
\end{figure}

%%%%%%%%%%%%%%%%%%%%%%%%%%%%%%%%%%%%%%%
Figure \ref{fig:fluxes} shows the different components of the energy flux (normalised by the injection rate) for the three cases. In case i), the total flux vanishes at $k_\perp<k_f$, while it is positive at $k_\perp>k_f$. At $k_\perp>k_f$, the flux of perpendicular kinetic energy is close to zero, and negligible compared to the large forward (positive) fluxes of parallel kinetic energy and potential energy. At the largest scales, all fluxes vanish, i.e. no energy is transferred to or from the large scales by nonlinear interactions. For intermediate scales between $k_\perp=k_f$ and $k_\perp\approx 5$, there is a wavenumber range over which there is a flux loop leading to zero net flux: the flux of pependicular kinetic energy is negative, i.e. inverse, while the kinetic energy in the parallel components of velocity and the potential energy show a positive (i.e. forward) flux, with the sum of the three cancelling out. In case ii), the flux loop persists at these intermediate scales, but the net flux is slightly negative (inverse), rather than zero. This inverse flux, which amounts to about $3\%$ of the energy injection rate, reaches all the way to the largest scales $k_\perp=1$, as the inset in figure \ref{fig:fluxes} shows. The parallel kinetic energy and potential energy fluxes are very similar to case i), being positive definite everywhere. In case iii), there is a strong net inverse flux, making up around $30\%$ of energy injection rate. Remarkably, while the dominant contribution to this inverse flux stems from the perpendicular kinetic energy, there is also an inverse flux of potential energy. In cases i) and ii), by constrast, the potential energy flux is positive definite. The strong stratification in case iii) breaks the passive-scalar-like evolution of the potential energy mentioned in section \ref{sec:theo_str}, which otherwise constrains the potential energy to cascade to small scales only. Moreover, the fact that both perpendicular kinetic energy and potential energy cascade inversely is compatible with the $\phi$ and $\omega_\|$ fields being linked by hydrostatic balance, which is shown to be the case in section \ref{sec:vis}.
%%%%%%%%%%%%%%%%%%%%%%%%%%%%%%%%%%%%%%%
\subsection{Well-resolvedness}
\begin{figure}
\includegraphics[width=0.329\textwidth]{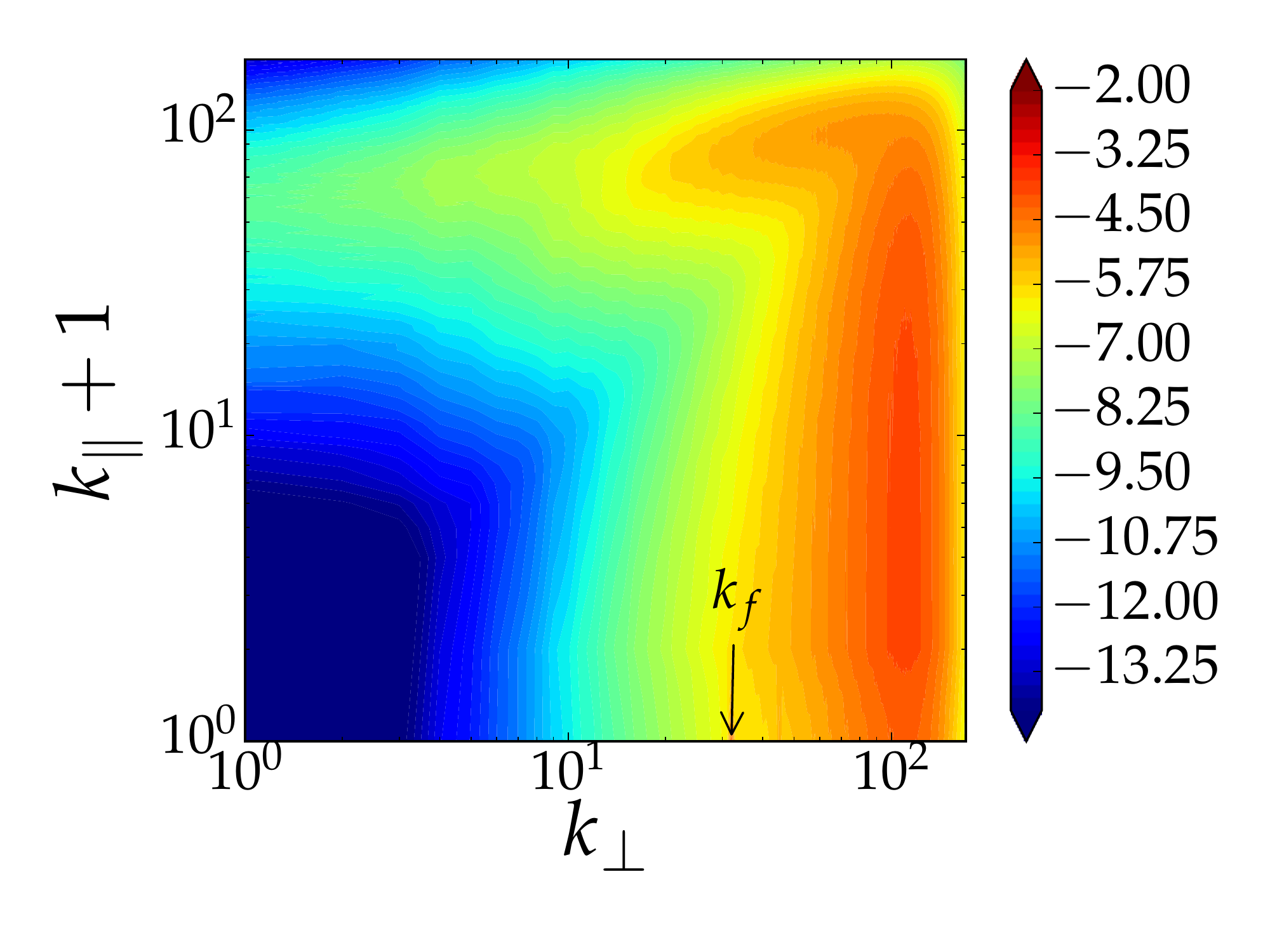}
\includegraphics[width=0.329\textwidth]{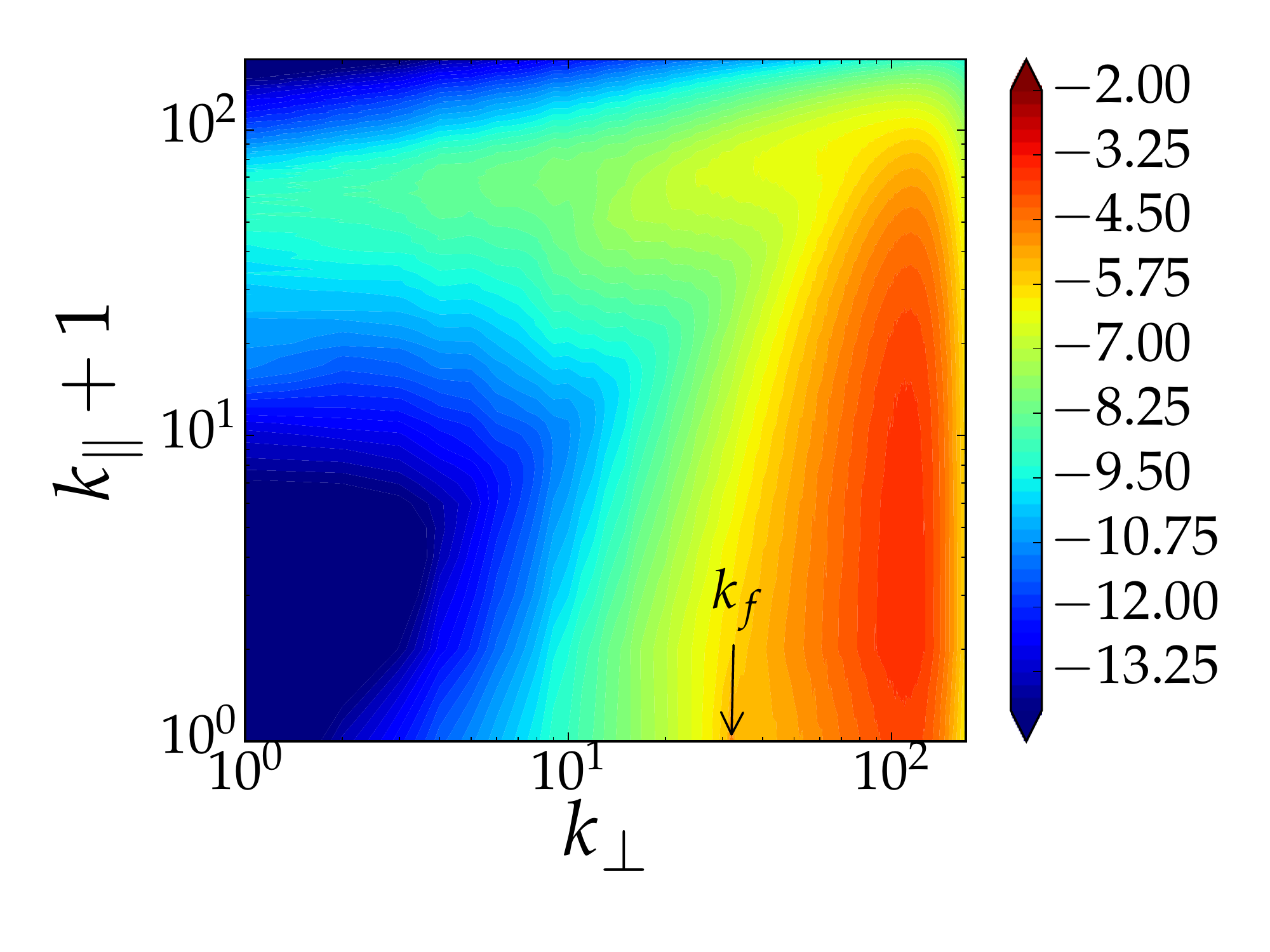}
\includegraphics[width=0.329\textwidth]{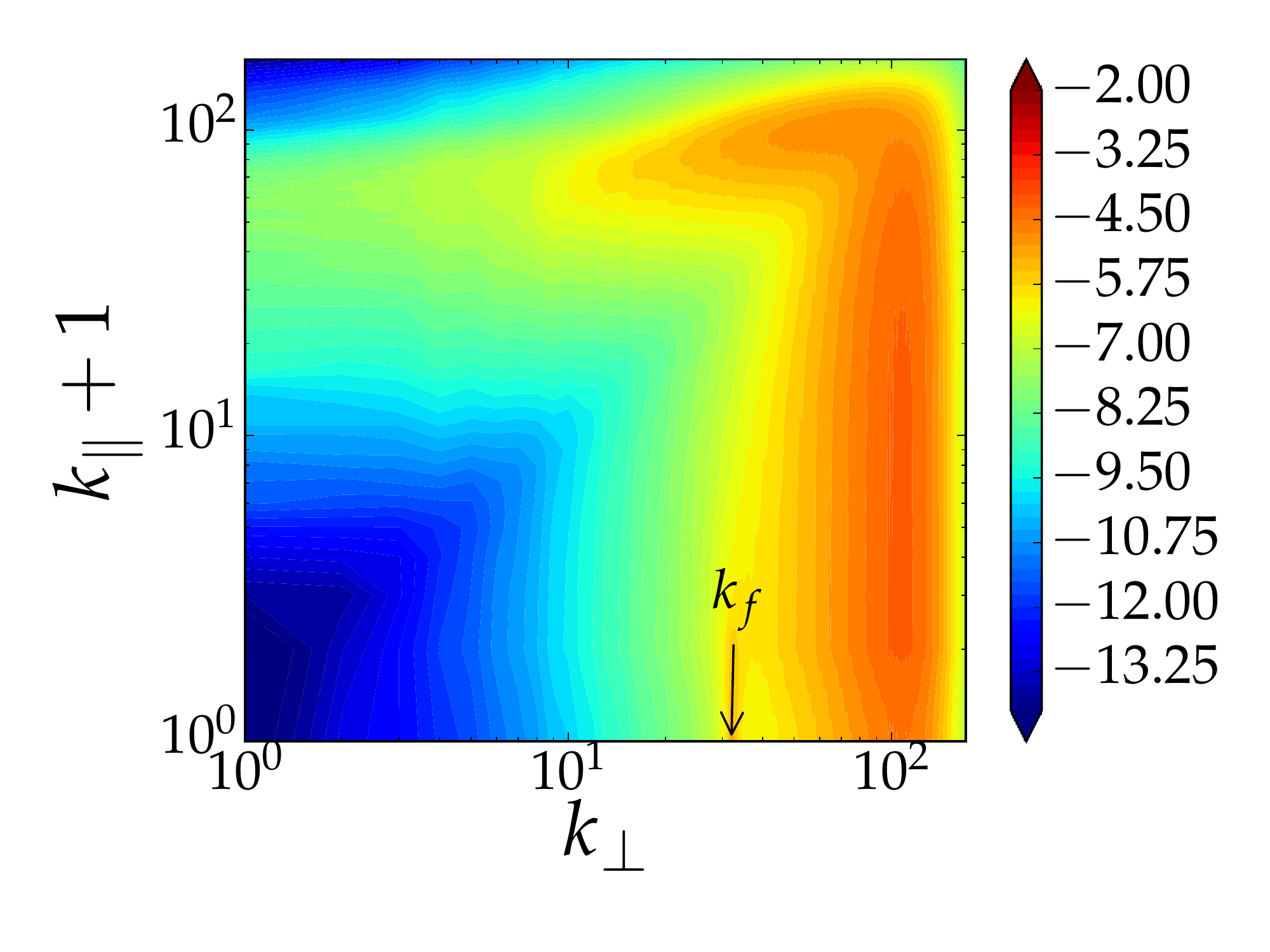}
\caption{Filled contour plots of 2-D dissipation spectra versus $k_\perp$, $k_\|$ for cases i) to iii) (left to right) 
%(case ii similar). 
The runs are well resolved since the maximum of dissipation is in the interior of the wavenumber domain.}
\label{fig:diss_spec}
\end{figure}

%%%%%%%%%%%%%%%%%%%%%%%%%%%%%%%%%%%%%%%
For each run, we verify well-resolvedness by inspecting the total dissipation spectrum $D_{tot}=D_{kin}+D_{pot}$ defined in eqs. (\ref{eq:diss_spec_kin}), (\ref{eq:diss_spec_pot}). For cases i) to iii), it is shown in figure \ref{fig:diss_spec}.
\BLUE{The integral over $k_\perp,k_\|$ of $D_{tot}$ expresses the total dissipation rate.  
The simulations are well-resolved if the maximum of the dissipation rate lies in the interior of the wavenumber domain
(as opposed to being found at the boundaries of the wavenumber domain).
For wavenumbers larger than the location of this maximum, the dissipation spectrum drops exponentially, implying exponential convergence: an increase of the resolution by a factor of $n$ will decrease the error due to spatial discretisation by a factor of $e^{-b n}$, for some positive $b$.    
Note that the presence of both vertical and horizontal viscosity/diffusivity is necessary for exponential convergence to exist. The maximum of $D_{tot}$ is clearly in the wavenumber domain in figures \ref{fig:diss_spec}.
This was also the case for all additional simulations at different Reynolds and Peclet numbers.} 

\BLUE{The fact that we do not examine higher values of $\lambda$, and smaller $\Fr$, in figure \ref{fig:overview} is due to the criterion of well-resolvedness described above}. At higher $\lambda$, the dissipation spectra showed significant dissipation at the largest $k_\|$ and the simulations were thus not well resolved. Therefore, these parameter values were not accessible at the present resolution. Simulations at higher resolution will be needed to confirm the tentative shape of the phase boundary between forward an inverse cascades at large $\lambda$ drawn in figure \ref{fig:ill}.

\BLUE{ Finally, besides examining numerical convergence of our simulations it is also important to examine
if our results are converged in $\Rey$,$\Pe$. For this we also repeated some of our runs in larger grid sizes doubling
$\Rey$ and $\Pe$, and verifying that the amplitude of the inverse flux did not change. Thus, up to the resolutions we were able to achieve, our results are robust.}

%%%%%%%%%%%%%%%%%%%%%%%%%%%%%%%%%%%
\subsection{Spatial structures}
\begin{figure}
\includegraphics[width=0.329\textwidth]{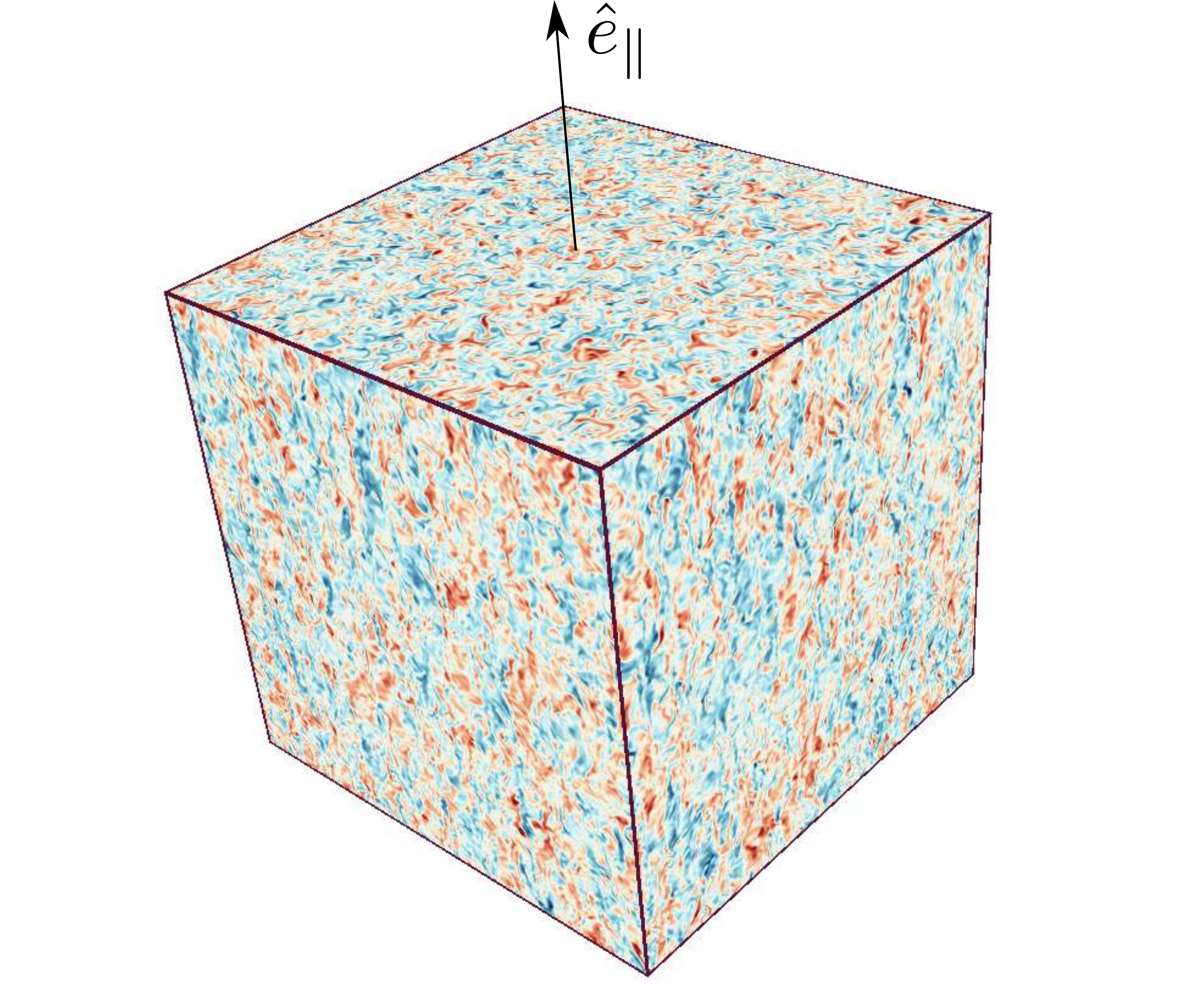}
\includegraphics[width=0.329\textwidth]{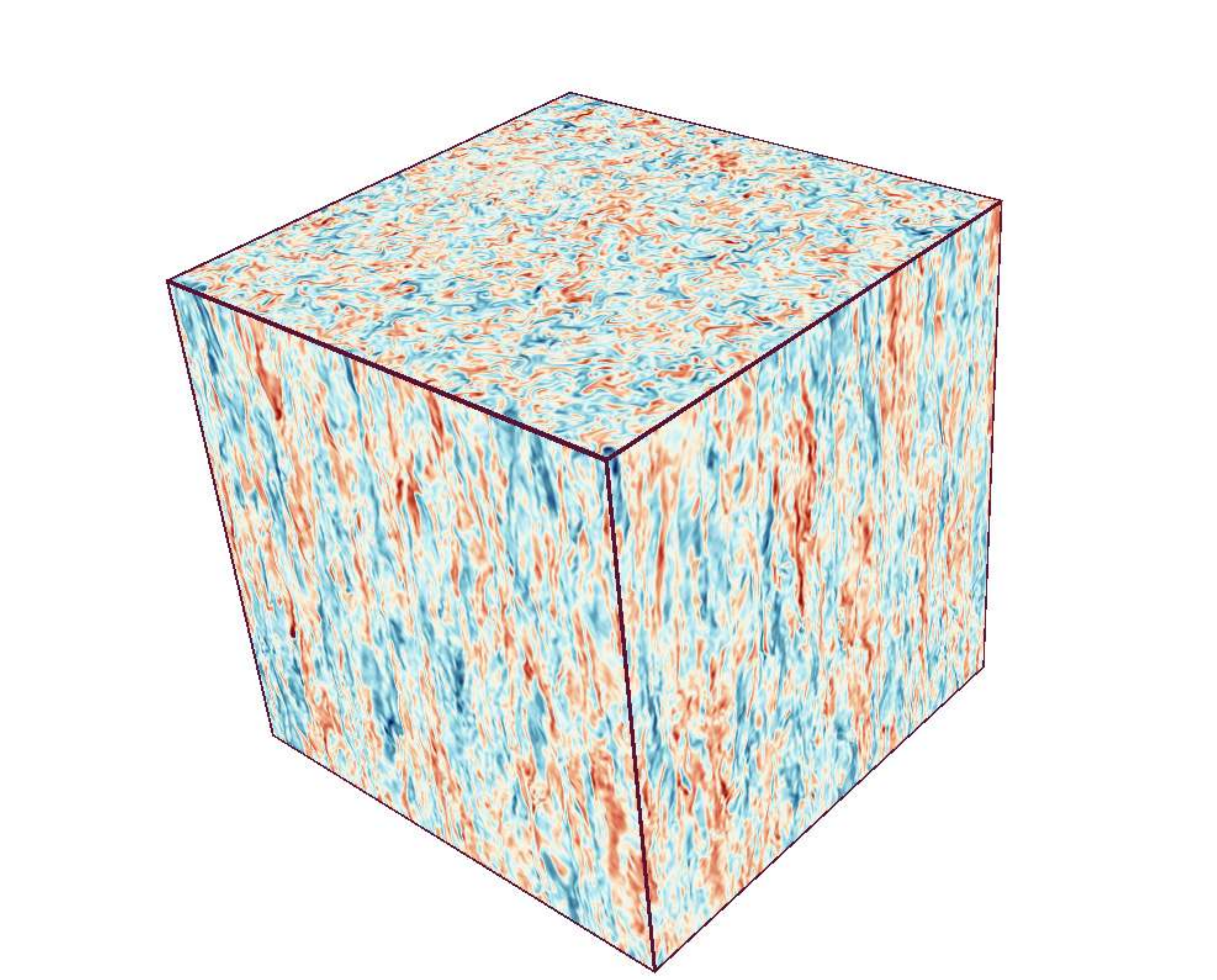}
\includegraphics[width=0.329\textwidth]{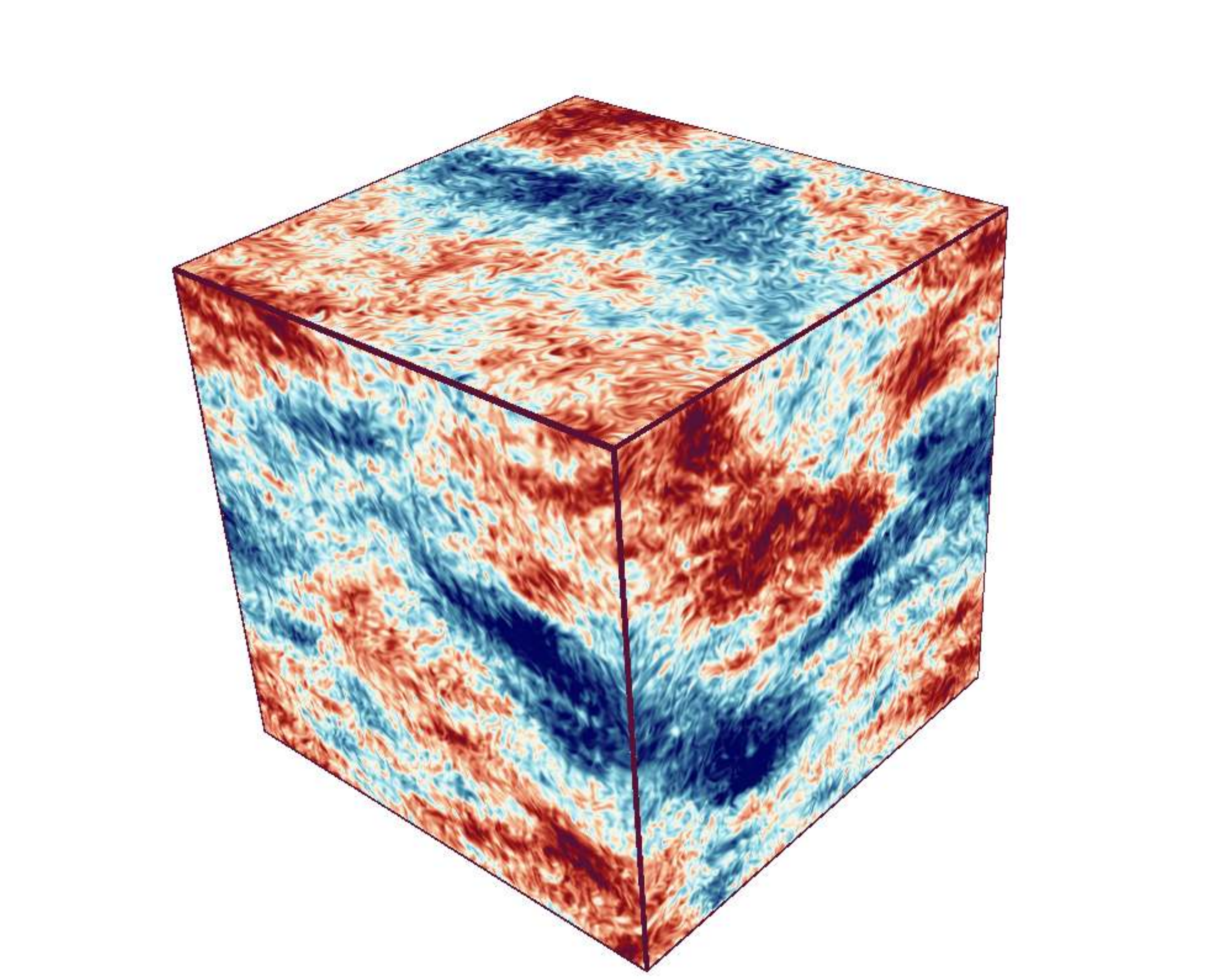}
\caption{Visualisation of the $\phi$ field. Left: case i, center: case ii, right: case iii. The black arrow indicates the parallel direction, it is the same for all other visualisations. The colour scale is the same in all three images, with blue colours representing negative values and red colours positive values. %($-3$ to $3$)
}
\label{fig:vis_b}
\end{figure}

\begin{figure}
\includegraphics[width=0.329\textwidth]{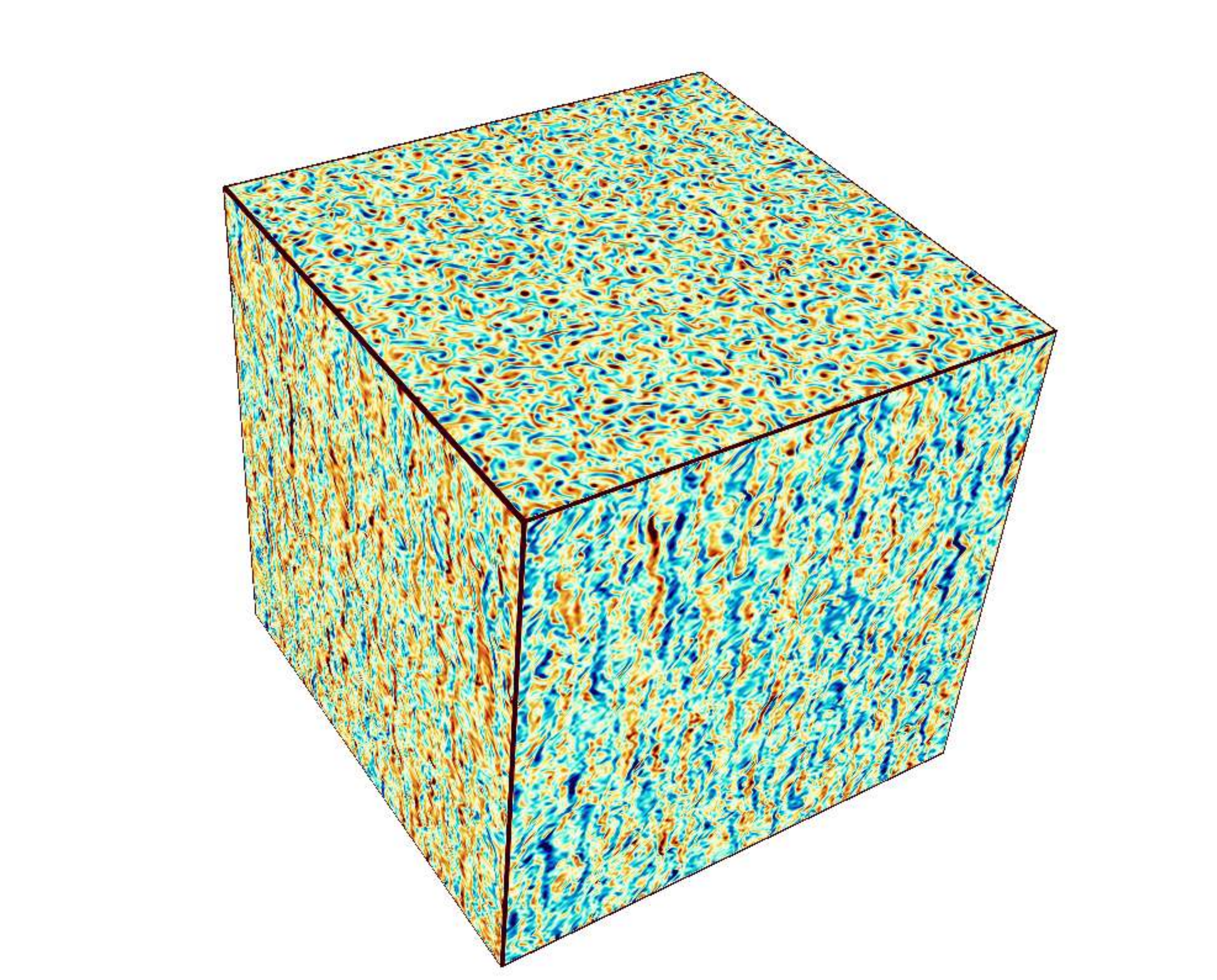}
\includegraphics[width=0.329\textwidth]{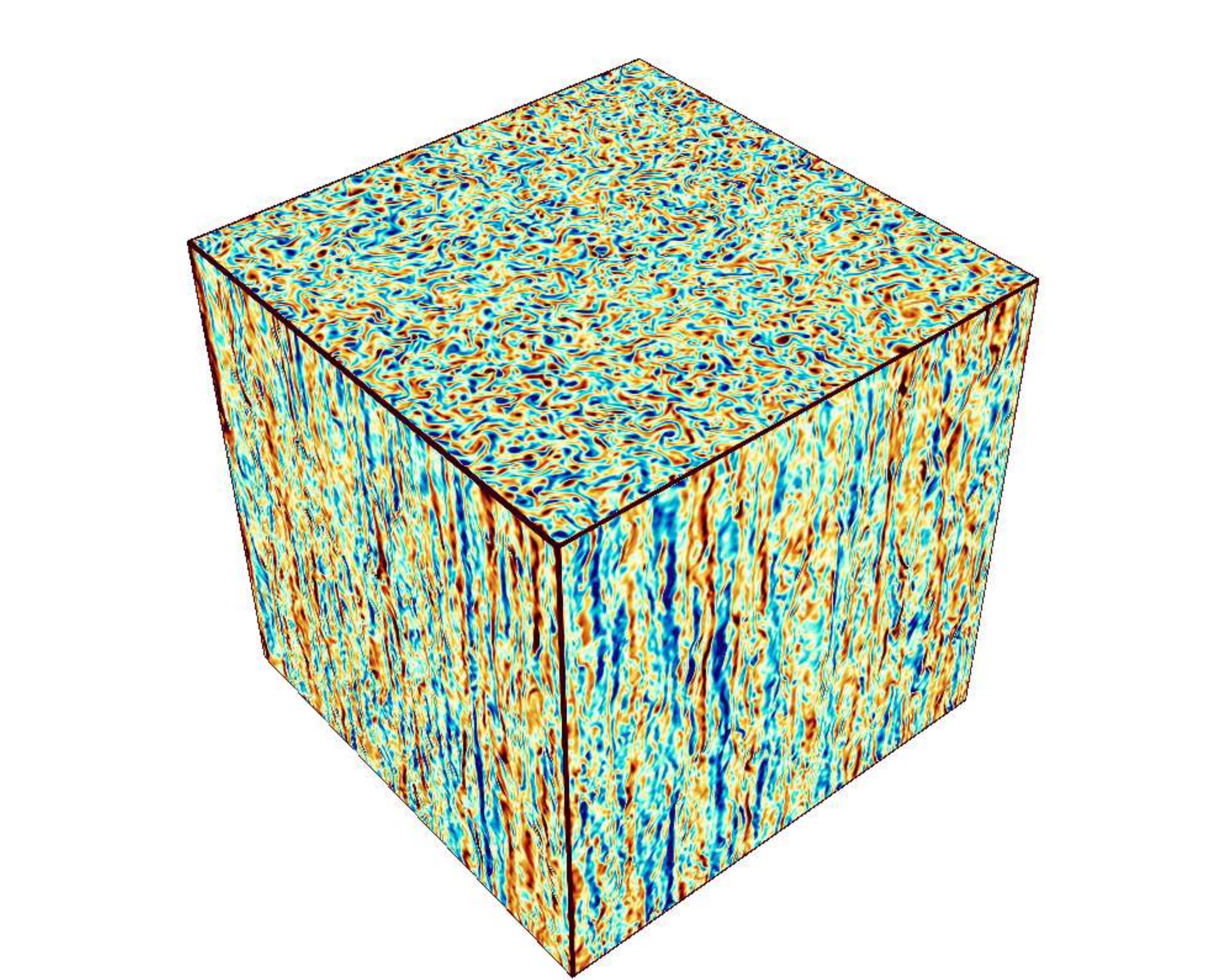}
\includegraphics[width=0.329\textwidth]{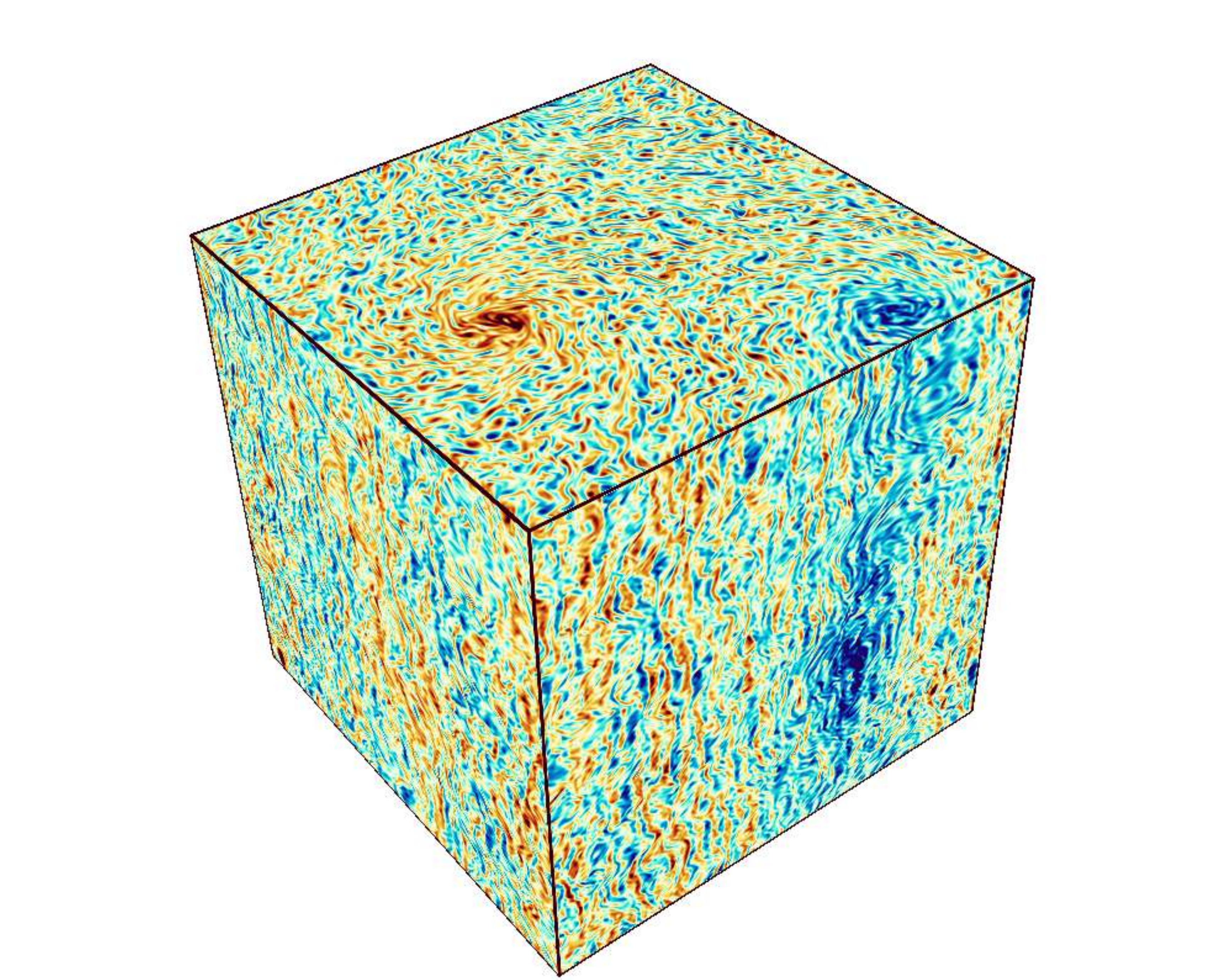}
\caption{Visualisation of the vorticity field. Left: case i, center: case ii, right: case iii.% (Colorbar range $-75,75$, color scheme Blue orange divergent). 
 The colour scale is the same in all three images, with blue colours representing negative values and red colours positive values. }
\label{fig:vis_w}
\end{figure}

%%%%%%%%%%%%%%%%%%%%%%%%%%%%%%%%%%%
\label{sec:vis}
Figure \ref{fig:vis_b} shows a visualisation of the density perturbation field $\phi$. For case i) there is large-scale organisation in the perpendicular direction, and there is some visible alignment in the parallel direction, in agreement with the 2-D spectra. In case ii), the rotation rate is stronger, leading to a more pronunced alignment in the vertical direction. However, the perpendicular scales in the $\phi$ field remain small. In case iii), the amplitude of the $\phi$ field is much higher than in cases i) and ii), and there is a clearly visible large-scale organisation in the parallel and perpendicular directions. In the parallel direction, there is a layering of density in approximately two layers, which is compatible with the 2-D potential energy spectra. In the perpendicular direction, one can see that the energy is at the largest scale $k_\perp=1$, since there is one large patch of positive $\phi$, and one of negative $\phi$ (periodic boundaries).\\
Figure \ref{fig:vis_w} shows a visualisation of the vorticity field. In case i), one sees no large-scale organisation in the perpendicular direction, and there is some rotation-induced alignment along the parallel direction. In case ii), the parallel alignment is more pronounced, since $\lambda$ is larger, equivalent to faster rotation. In the perpendicular direction, the condensation at the large scales has not yet proceeded far enough to be visible by eye, but 1-D spectrum in figure \ref{fig:1D_spectra} unequivocally shows that energy is piling up at large scales. Finally, in case iii), there is a clearly visible, high-amplitude pair of counter-rotating vortices on a small-scale background in the perpendicular direction. In the parallel direction, the alignment is weakened by the stronger stratification. We do not show visualisations of the parallel velocity field, since there it features only small-scale structures in all cases. \BLUE{We stress once more that in cases ii) and iii), what is shown is the transient state where the inverse cascade continues to develop, by contrast with the stationary state in case i).}

%%%%%%%%%%%%%%%%%%%%%%%%%%%%%%%%%%%%%%%%%%%%%%%%%%%%%%%%%%%%

\begin{figure}
\includegraphics[width=0.45\textwidth]{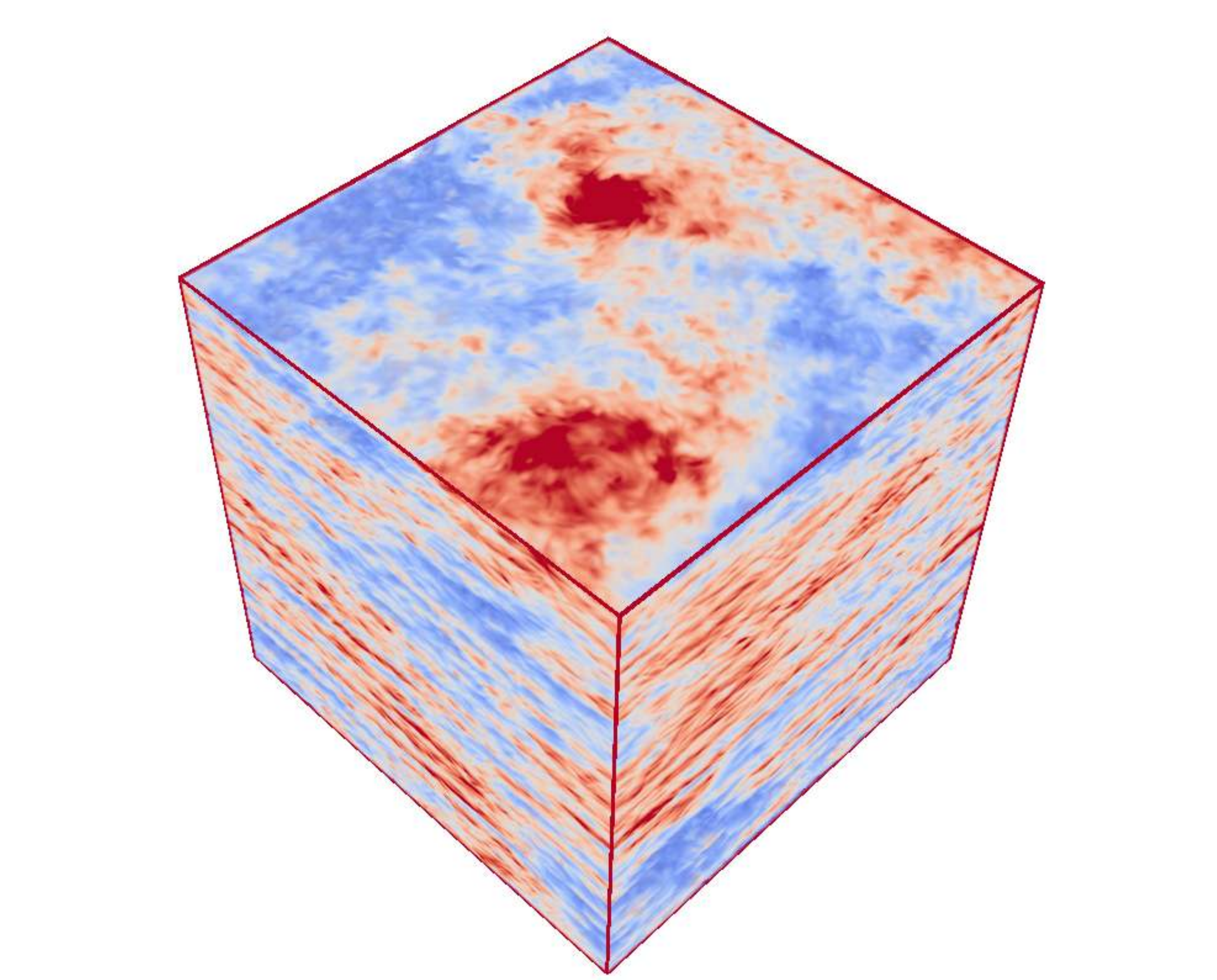}
\includegraphics[width=0.45\textwidth]{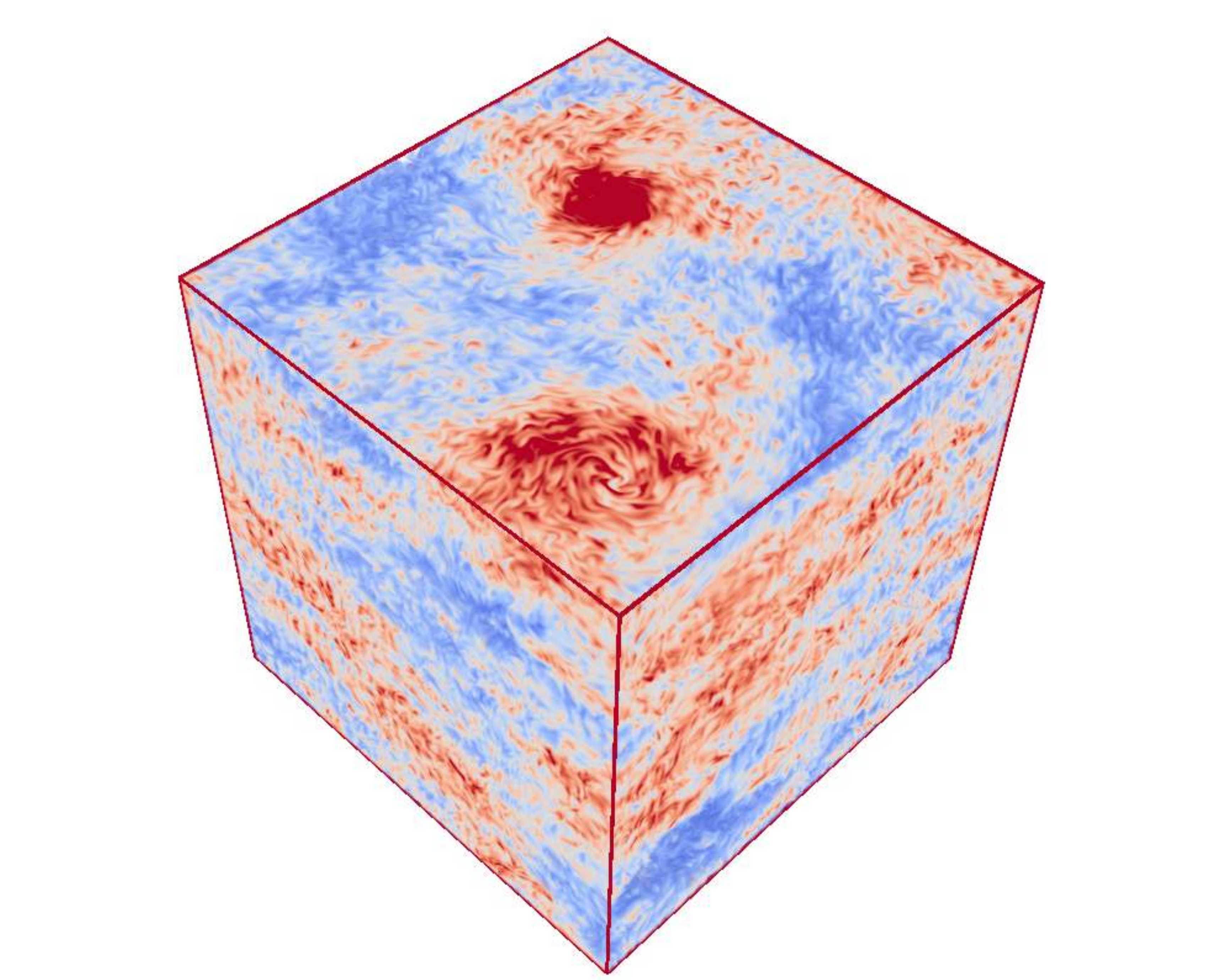}
\caption{Visualisation of the two terms involved in the hydrostatic balance relation (\ref{eq:hydrost_balance}). Red colours correspond to positive values and blue colours to negative values. The left panel shows $2\lambda \partial_\| \psi$, the right panel $-\phi/\Fr$. The two fields are clearly correlated.}
\label{fig:check_hb}
\end{figure}

Figure \ref{fig:check_hb} shows visualisations of the two terms involved in hydrostatic balance (\ref{eq:hydrost_balance}): parallel pressure gradient $2\lambda \partial_\| \psi$ and the buoyancy force $-\phi/\Fr$. The two fields are visibly highly correlated. Together with the spectra and fluxes above, this validates the proposed explanation of the phenomenology of case iii) based on hydrostatic balance \BLUE{in the quasi-geostrophic limit}.

%%%%%%%%%%%%%%%%%%%%%%%%%%%%%%%%%%%%%
%%%%%%%%%%%%%%%%%%%%%%%%%%%%%%%%%%%%%
\section{Discussion}              %%%
\label{sec:discussion}            %%%
%%%%%%%%%%%%%%%%%%%%%%%%%%%%%%%%%%%%%
%%%%%%%%%%%%%%%%%%%%%%%%%%%%%%%%%%%%%

\BLUE{
In this paper we investigated energy cascades in stably stratified, rapidly rotating turbulence within an elongated domain. Using a large number of numerical simulations of a reduced system, we constructed a phase diagram of the system marking the regions in phase-space where inverse cascade is met. Two different behaviors were noted.
First, for weak stratification, an inverse cascade appears above a threshold $\lambda_c$ that is an increasing function of $\Fr^{-1}$. For $\Fr^{-1}=0$ it recovers the non-rotating threshold. In this limit, the energy of the in-plane velocity components $\bf u_\perp$ cascades inversely while potential energy and kinetic energy related to $u_\|$ cascade forward. 
%both for $\Fr \gtrsim 1$ and for large $\lambda=(\Ro h)^{-1}$, while a forward energy cascade is observed for $\Fr^{-1}\lesssim 1$ and $\lambda$ below a threshold $\lambda_c$. 
For strong stratification, inverse cascades appears for $\Fr^{-1}\lesssim \Fr_c^{-1}$
where this second threshold $\Fr_c$ is independent of $\lambda$ and $\Fr_c\simeq 1$.
In this limit we found that approximate hydrostatic balance holds, leading to a non-trivial inverse cascade of both potential and kinetic energy. 
%Future research should aim to characterise in more detail the approach to hydrostasy in this extreme parameter regime, in order to better assess its impact on the energy cascades in rotating and stratified turbulence within elongated domains.
}

\BLUE{
Our approach was based on asymptotic reduction, allowing us to reliably achieve the parameter regime of interest at comparatively moderate numerical cost. 
The validity of this approximation and its limitations, however, need to be discussed.
We remind the reader that in our approach the limits $\Ro\to0$ and $h\to\infty$ are taken while keeping the product 
$\lambda^{-1}=\Ro h$ and all other parameters $\Rey,\Pe,\Fr,L/\ell_{in}$ fixed.
%
%The asymptotics used here are valid for $\Ro\ll1$, and $\Rey,\Pe,\Fr\ll\Ro^{-1}$, as well as $\Rey,\Pe,\Fr\ll H/\ell_{in}$, while the horizontal box size $L\ll H$. 
First we would like to comment that with this limiting procedure weak wave turbulence is not met in our simulations. Weak turbulence requires taking the tall-box limit $h\to\infty$ first and then $\Ro\to0$ , so that $\lambda=1/h\Ro\to0$. Weak wave turbulence (for $\Fr^{-1}=0$) predicts only forward cascade \citep{galtier2003weak} and this is indeed what we find for $\lambda\to0$. 
Thus, the two regimes (the present asymptotic result and weak wave turbulence prediction) appear to commute for the weak stratification limit. For $\Fr^{-1}>0$ up to our knowledge 
there is no theoretical result. It is of particular interest to know if the inverse casca  observed
in the present limit for $Fr^{-1}>1$ persists 
or not as $\lambda$ is decreased below the range of validity  of the present approximation and into the rotating and stratified wave turbulence regime. If not, this would imply that the shape of the phase boundary will change as smaller values of $\lambda$ (or order $\mathcal{O}(\epsilon)$) are approached. This needs to be investigated by future theoretical work and numerical simulations of the full Boussinesq equations.  }

\BLUE{A second issue that needs to be discussed is whether the limits $\Ro^{-1}, h\to\infty $ and $\Rey,\Pe\to\infty$ also commute. Generally, one is interested in the large-Reynolds-number and large-Peclet-number limits. The energy fluxes obtained upon taking these limits first, and then taking $\Ro^{-1},h\to \infty$ will not necessarily give the same result as when the order is reversed.  A particular limitation of the asymptotically reduced equations is that the perpendicular motions are required to be geostrophically balanced. In the full system at large Reynolds numbers, this balance may be broken at the small scales for which isotropy might be restored. This could alter the energy transfer properties of the system. In particular, it is known that the presence of stratification leads to smaller and smaller vertical scales \citep{Billant2001self} that have been argued to hinder the inverse cascade. 
However, we need to note that the scale at which geostrophic balance is broken becomes smaller
and smaller as $\Ro$ is decreased, so that for sufficiently small $\Ro$ the separation between the inversely-cascading geostrophically-balanced scales and the forward cascading isotropic scales will increase and the interaction between the two scales will become weak.  }
\BLUE{ 
Finally we need to also discuss the limit of large $\Lambda=L/\ell_{in}$.
If an inverse cascade is present in the horizontal plane larger and larger
horizontal scales are reached. When these scales become of the order $1/\epsilon$
the present approximation also ceases to be valid for these scales. 
All these limitations call 
for investigating in the future, also at finite values of the parameters 
using the full rotating and stratified Navier Stokes equations.
}

\BLUE{
Concluding, we would like to note that for the purely rotating problem, \citep{di2020phase} undertook a step in this direction, showing that meta-stable vortex-crystal states appear near the transition to an inverse cascade, while such states were not seen in the reduced equations. It is therefore a possibility that the complete phase diagram of rapidly rotating and stratified turbulence is more complex than anticipated.}

%%%%%%%%%%%%%%%%%%%%%%%%%%%%%%%%%%%%%%%%%%%%%%%%%%%%5
\backsection[Acknowledgements]{We thank Basile Gallet for his critical assessment of our manuscript, Santiago Benavides and Pablo Mininni for fruitful discussions, and two anonymous referees for their helpful comments and suggestions. }

\backsection[Funding]{This work was granted access to the HPC resources of MesoPSL financed by the Region Ile de France and the project Equip@Meso (reference ANR-10-EQPX-29-01) of the programme Investissements d'Avenir supervised by the Agence Nationale pour la Recherche and the HPC resources of GENCI-TGCC \& GENCI-CINES (Project No. A0070506421, A0080511423, A0090506421). This work has also been supported by the Agence nationale de la recherche (ANR DYSTURB project No. ANR-17-CE30-0004). AvK acknowledges support by Studienstiftung des deutschen Volkes.}

\backsection[Declaration of interests]{{\bf Declaration of Interests}. The authors report no conflict of interest.}

\backsection[Author ORCID]{A. van Kan https://orcid.org/0000-0002-1217-3609; A. Alexakis https://orcid.org/0000-0003-2021-7728}

%\appendix

%\section{}\label{appA}
% This appendix contains sample equations in the JFM style. Please refer to the {\LaTeX} source file for examples of how to display such equations in your manuscript.

%\bibliographystyle{jfm}
%\bibliography{jfm}
%Use of the above commands will create a bibliography using the .bib file. Shown below is a bibliography built from individual items.

\bibliographystyle{jfm}
\bibliography{biblio}

%% End of file `jfm2esam.bib'.

\end{document}